\documentclass[final,1p,times]{elsarticle}

\usepackage{lineno,hyperref}
\usepackage[utf8]{inputenc}
\usepackage[english]{babel}
\usepackage[T1]{fontenc}
\usepackage{pdfwidgets}
\usepackage{natbib}
\usepackage{graphicx}
\usepackage{txfonts}
\usepackage{mathptmx}
\usepackage{epsfig}
\usepackage{amssymb}
\usepackage{amsmath}
\usepackage{verbatim}
\usepackage{caption}
\usepackage{subcaption}
\usepackage{amsmath}
\modulolinenumbers[5]

\journal{Astroparticle Physics}

\begin{document}

\begin{frontmatter}

\title{Acoustic Neutrino Detection In a Adriatic Multidisciplinary Observatory (ANDIAMO)}


\author[mymainaddress]{Antonio Marinelli}
\ead{antonio.marinelli@na.infn.it}

\author[mymainaddress]{Pasquale Migliozzi}
\ead{pasquale.migliozzi@na.infn.it}

\author[mymainaddress]{Andreino Simonelli}
\ead{andreino.simonelli@na.infn.it}


\address[mymainaddress]{INFN - Sezione di Napoli, Complesso Univ. Monte S. Angelo, I-80126 Napoli, Italy}

\begin{abstract}
The existence of cosmic accelerators able to emit charged particles up to EeV energies has been confirmed by the observations made in the last years by experiments such as Auger and Telescope Array. The interaction of such energetic cosmic-rays with gas or low energy photons, surrounding the astrophysical sources or present in the intergalactic medium, guarantee an ultra-high-energy neutrino related emission. When these energetic neutrinos interact in a medium produce a thermo-acoustic process where the energy of generated particle cascades can be conveyed in a pressure pulse propagating into the same medium. The kilometric attenuation length as well as the well-defined shape of the expected pulse suggest a large-area-undersea-array of acoustic sensors as an ideal observatory. For this scope, we propose to exploit the existing and no more operative offshore (oil rigs) powered platforms in the Adriatic sea as the main infrastructure to build an acoustic submarine array of dedicated hydrophones covering a surface area up to 10000 Km$^{2}$ and a volume up to 500 Km$^3$. In this work we describe the advantages of this detector concept using a ray tracing technique as well as the scientific goals linked to the challenging purpose of observing for the first time ultra-high-energy cosmic neutrinos. This observatory will be complementary to the dedicated radio array detectors with the advantages of avoiding any possible thermo-acoustic noise from the atmospheric muons.
\end{abstract}


\end{frontmatter}


\section{Introduction}
\begin{figure}[h!]
\centering
\includegraphics[scale=0.5]{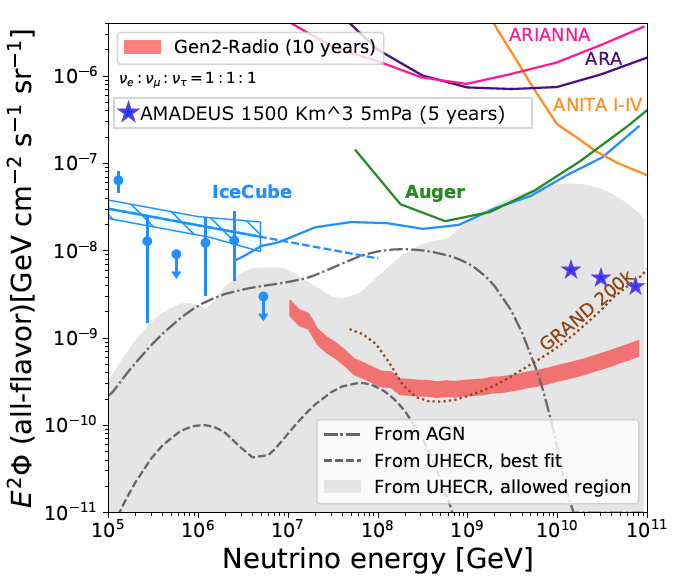}
\caption{In this plot are reported the expected sensitivities of the main radio array detectors for UHE neutrinos, comprising the future IceCube Gen2-Radio~\citep{IceCube-Gen2:2020qha}, in comparison with the capabilities of an acoustic array and the main UHE diffuse neutrino expected signals. In details, with the blu stars it is reported the sensitivity of one of the few large scale undersea acoustic detectors recently simulated, in this case by the AMADEUS group~\citep{2017EPJWC.13506002S}. This simulated acoustic array refers to a huge sparse array (1500 Km$^{3}$) of sensors with a density of $\sim100$ per Km$^{3}$ and a kinetic threshold of 5 mPa.}
\label{fig:sensy}
\end{figure}

The ultra-high-energy cosmic-rays (UHECRs) observed by Pierre Auger apparatus~\cite{Abraham:2010mj} and Telescope Array~\cite{AbuZayyad:2012ru} suggest the presence of astrophysical accelerators who can originate EeV neutrinos considering the possible cosmic-rays (CRs) interaction~\cite{Halzen:2002pg} with gas or photons in the source environment~\cite{Bahcall:1999yr} or along the path to the Earth~\cite{Kotera:2010yn}. 
On the other hand, the astrophysical flux observed by IceCube telescope during the last decade~\cite{Aartsen:2014gkd,Aartsen:2015rwa} demonstrates that the former interactions occur with CRs accelerated at least up to an energy of hundred of PeVs~\cite{Aartsen:2013jdh}. Who are the main responsibles for these neutrino emissions is still a matter of debate. However, while PeV-EeV neutrinos can hardly be produced  in our Galaxy we can expect the extragalactic accelerators to be the main candidates for that emission.
The muonic neutrino at 290 TeV observed by IceCube on 22 of September 2017 from a direction compatible with the blazar TXS 0506+056, in coincidence with gamma-ray flare observed by Fermi-LAT and MAGIC telescopes, makes blazars to climb the charts of the very-high-energy neutrino emitter candidates. 
Most of the hadronic models related to this class of powerful sources suggest a favourite range of neutrino production between several hundreds of TeV up to a few EeVs. This implies that the astrophysical samples of IceCube events can be only partially explained by the blazar emission. This hypothesis can be confirmed whenever we will have ultra-high-energy (UHE) neutrino observations.
Another important neutrino emitter candidate who is expected to emit up to EeV energies is represented by the gamma-ray burst (GRB) who are able to accelerate CRs up to these energies~\cite{Murase:2005hy} during the so-called prompt phase~\citep{Biehl:2017zlw}.
In addition to the UHE neutrinos that could arrive directly from one of the mentioned astrophysical sources, at energies above PeV, there is also the cosmogenic neutrino flux produced by the interaction UHECRs  with the astrophysical background radiation fields. This photon target is represented by the extra-galactic background light (EBL) and the cosmic microwave backgrounds (CMB) along the path between the source and the Earth~\cite{Beresinsky:1969qj}. While the former signal is expected to produce a spotted observation in correlation with UHE sources, the latter is expected to be a uniform signal over the full sky. The possibility to identify single astrophysical sources of UHE neutrinos can be challenging for such a detector due to the number of events needed. The information about the astrophysical spectral shape observed will be fundamental to identify a class of astrophysical emitters in this range of energy.\\
The detection of neutrinos from PeV to EeV range could represent a major breakthrough and go beyond the capabilities of Cherenkov neutrino telescopes like IceCube~\citep{IceCube:2016zyt}, Antares~\citep{ANTARES:2011hfw}, Baikal~\citep{Aynutdinov:2012zz} and KM3NeT~\citep{KM3Net:2016zxf}. A larger instrumented surface and a different approach to collect the neutrino debris signature are needed. The two main techniques proposed for the neutrino detection at UHEs are the radio detection in atmosphere and  the acoustic detection in water/ice thanks to the Askarian effect~\cite{Askaryan:1962hbi} happening when high-energy ionizing particles pass through a dense medium. The radio detection techniques improved during the last decades thanks to the installation of several prototypes and telescopes as the pioneering RICE~\cite{Kravchenko:2001id,Kravchenko:2011im}, ARA~\cite{Allison:2015eky} , ARIANNA~\cite{Barwick:2014rca,Barwick:2014pca}, RNO-G~\cite{Aguilar:2020xnc}, GRAND~\cite{2020SCPMA..6319501A} as well as the balloon experiments like ANITA~\cite{Gorham:2008dv}. 
Even if up-to-date none of these radio arrays were able to find an excess of signal statistically significant as recently reported by ARA collaboration~\cite{Allison:2019xtn} the analyses done helped to better understand the time dependent behaviour of the radio arrays as well as to improve the rejection of noise. \\
The use of acoustic techniques for UHE neutrino telescope favours the exploiting of natural water/ice reservoirs like seas and big lakes~\citep{Kurahashi:2010ei,Lehtinen2002}. Thanks to the accelerated expansion of the cylindrical volume heated by the interaction of neutrino with a nucleon and the production of a hadronic cascade~\cite{Askarian:1979zs,Learned:1978iv} an acoustic array can reconstruct the generating neutrino event. The two experimental setups represents a complementary approach to cover a larger solid angle, while the first favours the Earth-skimming neutrino events, the second privileges the down-going events.\\
Despite the test setups done with small acoustic arrays like SPATS~\cite{Abdou:2011cy}, O$\nu$DE5\cite{Riccobene:2009zz}, ACoRNE~\cite{Danaher:2007zz}, AMADEUS~\cite{Aguilar:2010ac} and SOUND~\cite{Vandenbroucke:2004gv} already installed for monitoring activities or built as sub-detector of major Cherenkov telescopes, at different latitudes, a large scale underwater/ice acoustic telescope has not yet finalized.
For a dedicated acoustic array, a large area coverage $\mathcal{O}(1000~$Km$^{2})$ and a correspondent sizeable volume $\mathcal{O}(100~$Km$^{3})$~\cite{Kurahashi:2010ei} are needed, even considering a sparse units distribution.\\
In this work we introduce the possibility of exploiting the ENI not operative powered oil rigs in the offshore of the Adriatic sea. The oil upstream activity is terminated but the infrastructures are still available for scientific purposes.
We show that the Mediterranean Sea certainly represents a preferential environment for an acoustic telescope due to the effects of water temperature, currents, and salinity~\cite{Aynutdinov:2012zz}. 
Moreover, we prove that by exploiting the shape and the propagation pattern of the acoustic wave-field generated by UHE neutrino cascade we can cover an array of unprecedented area.
This description can be translated to a large-scale distribution of small sized strings equipped with equidistant acoustic sensors.
In Fig.~\ref{fig:sensy} we report the possible sensitivity of an extended acoustic array simulation~\citep{2017EPJWC.13506002S}, made by the AMADEUS group, in comparison with the sensitivity of mentioned radio-arrays and the major expected diffuse UHE neutrino signals.
It is important to say that the acoustic array proposed here, ANDIAMO (Acoustic Neutrino Detection in a Multidisciplinary Observatory), can be used not only for UHE neutrino detection but also for marine biology and geophysical studies.
The large amount of data that will be collected over a wide span of frequencies can be used in full by many scientific communities.
Monitoring the oceanographic parameters of a large portion of the Adriatic sea is an effective way to study climate changes, and the extremely high sensitivity of the acoustic sensors is suitable for seismic monitoring of local and teleseismic events.
The shallow waters of the site permits the installation of ocean bottom seismometer (OBS) platforms at low cost.
All the equipment installed for astroparticle physics purposes is needed for a better understanding of the Sea in terms of medium where our expected signal propagates. At the same time, the collected signals will serve a vast branch of science with a large amount of data.

\section{The Science case}
The study of possible neutrino emission beyond the energies already observed by IceCube telescope can answer important questions about the processes happening on astrophysical accelerators, as well as the nature of local Dark Matter (DM) scenarios. Even though the identifications of single UHE accelerators, like the ones reported in Fig.~\ref{fig:sensy1} and Fig.~\ref{fig:sensy2} can be challenging, the observation of that kind of neutrino signal and the estimation of an isotropic flux represent in itself a major step for the astroparticle community.

\subsection{Cosmogenic neutrinos}
The ankle observed in the cosmic-ray spectral energy distribution at around $5\times10^{18}$ eV is interpreted as the transition point between the Galactic and the extragalactic origin of the observed comic rays~\citep{Berezinsky:2002nc,Aloisio:2012ba}.
With the observations made by the Pierre Auger experiment~\cite{Abraham:2010mj} and Telescope Array~\cite{AbuZayyad:2012ru} we already collected hundreds of these events with energy above $10^{18}$ eV. When these events interact with background photons, like CMB~\citep{deBernardis:2000sbo} and EBL~\citep{Franceschini:2008tp}, they are producing gamma-rays and neutrinos through the decay of neutral and charged pions ($\pi^{0}\rightarrow \gamma\gamma$,~$\pi^{+}\rightarrow e^{+}\nu_{e}\nu_{\mu}\overline{\nu}_{\mu}$) who are produced trough the $\Delta^{+}$ channel as described in the following equation: 
\begin{equation}
p+\gamma_{bg} \rightarrow \Delta^{+} \rightarrow 
\begin{cases}
& p + \pi^{0},\\
& n + \pi^{+}
\end{cases}
\label{proton_int}
\end{equation}
The energy threshold of this process is of 1.08 GeV in the reference system of the interacting particles~\citep{Mucke:1998mk}. This is also the main reason of the distortion of the proton spectrum above $3\times 10^{19}$ eV during propagation, known as the Greisen–Zatsepin–Kuzmin cutoff (\citep{Greisen:1966jv}, \citep{Zatsepin:1966jv}).\\
Considering that the CMB density increases with the distance as $(1+z)^{3}$ we should expect also the cosmogenic neutrino production depending on the redshift considered. An equivalent dependency should be expected for UHE neutrino produced by the interaction of cosmic rays with EBL event though the EBL parameters are less known in respect to the CMB, due to the spectral evolution of optical, infrared and ultraviolet. In this case, being the EBL photons more energetic that the CMB ones, the energy threshold for neutrino production through the photopion process becomes lower.\\
Beta-decay can contribute to the cosmogenic neutrino production as well, due to the decay of neutron obtained with the charged pion production:
\begin{equation}
n \rightarrow p + e^{-} + \overline{\nu}_{e}
\end{equation}
Heavier nuclei with $Z>1$ can also produce beta-decay process, adding a possible contribution to the UHE neutrino flux generated through photopion interaction. 
Following the presented description, the cosmogenic neutrinos originated by the interaction of UHECRs with thermal photons are expected to follow a double peaks spectral energy distribution (SED) due to the targets considered. The low energy peak, at around $10^{16}$ eV, can be associated to the interaction of cosmic-rays with EBL and the neutron beta-decay. Conversely, the high energy peak, is expected to be the result of cosmic-rays interaction with the CMB photons~\citep{Das:2020nvx}.
On the other hand the multi-messenger studies of UHECRs and diffuse gamma-rays cannot constrain the expected cosmogenic neutrino spectrum~\citep{Ackermann:2014usa}.

\subsection{Extragalactic accelerators}
\begin{figure}[h!]
\vspace{-2cm}
\centering
\includegraphics[scale=0.45, angle =270]{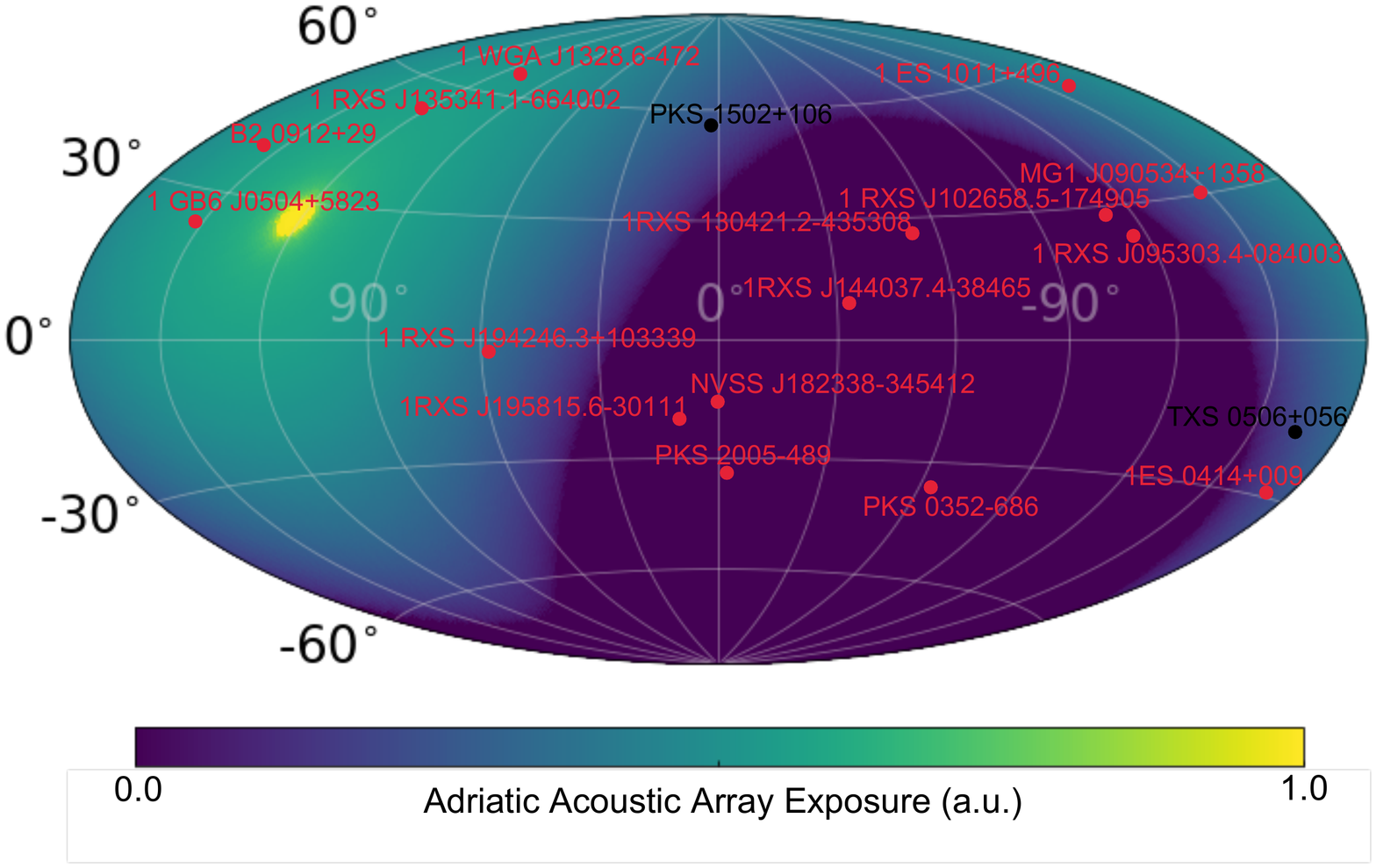}
\vspace{-1cm}
\caption{Exposure Skymap (in Galactic coordinates) of the presented UHE neutrino telescope. In the Skymap are reported also the sources of the 2FHL catalog spatially correlated with a UHECR observed by Auger or Telescope Array (within $5^{\circ}$). In Black are reported the blazars TXS0506+056 and PKS1502+106 as the two main neutrino emitter candidates observed up to now. A maximum angular acceptance of 45$^{\circ}$ from the local zenith is considered.}
\label{fig:sensy1}
\end{figure}

\begin{figure}[h!]
\centering
\includegraphics[scale=0.45, angle=270]{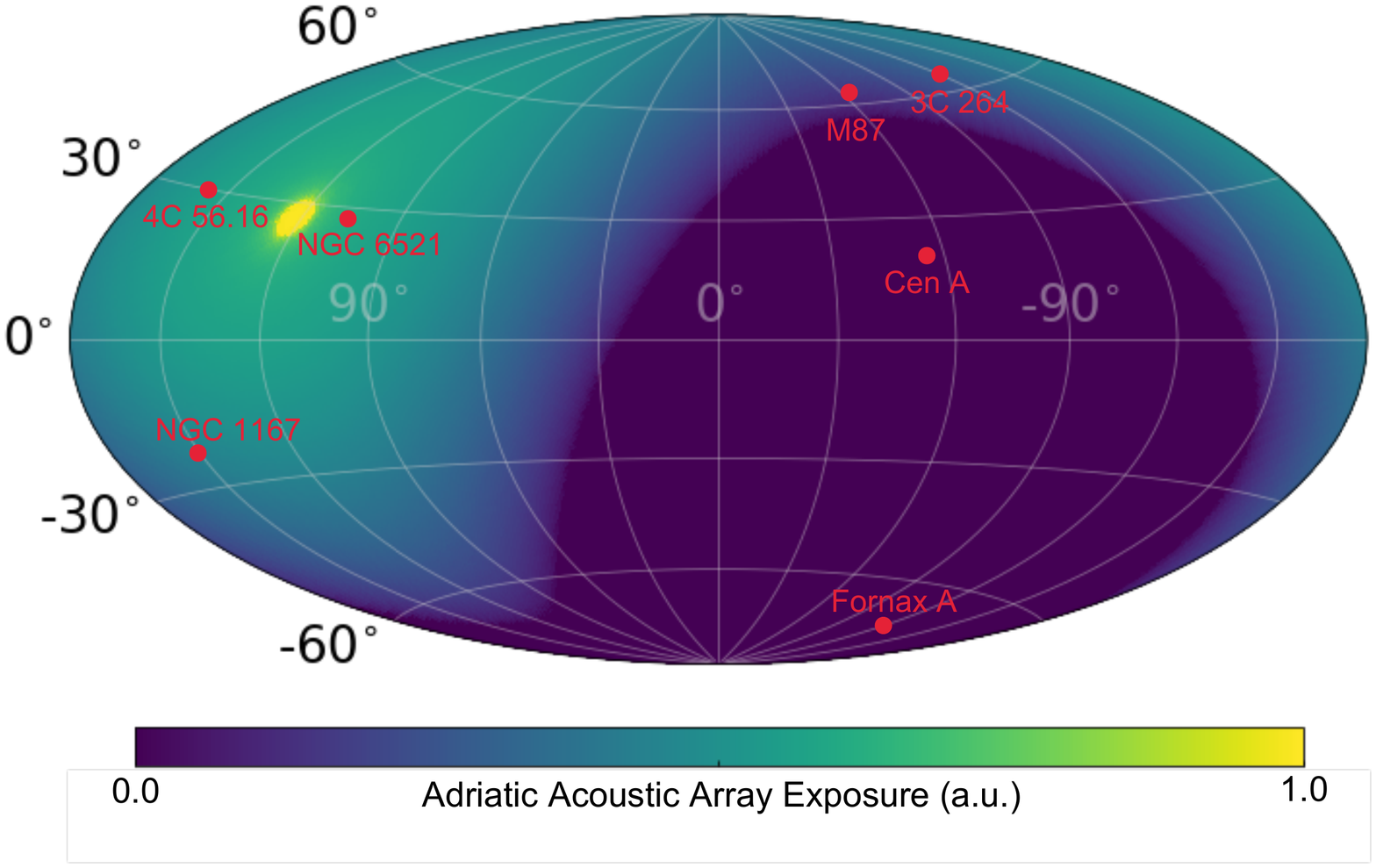}
\vspace{-1cm}
\caption{Exposure Skymap (in Galactic coordinates) of the presented UHE neutrino telescope. In the Skymap also the radio galaxies, with a radio luminosity $\nu L_{\nu} > 2 \times 10^{40}$ erg s$^{-1}$,  expected to have the possibilities of accelerates cosmic-rays up to UHE in the giant lobes~\citep{Matthews:2018rpe}.
A maximum angular acceptance of 45$^{\circ}$ from the local zenith is considered.}
\label{fig:sensy2}
\end{figure}

As mentioned in the previous subsection, the acceleration of cosmic-rays above $10^{18}$ eV can occur in astrophysical plasmas when large-scale motion, such as shocks and turbulent flows, is transferred to individual charged particles. The maximum energy of accelerated particles, $E_{max}$, can be estimated by requiring that the gyroradius of the particle be contained in the acceleration region. Therefore, for a given strength, B, and coherence length, L, of the magnetic field embedded in an astrophysical region, $E_{max} = Z \cdot e \cdot B \cdot L$, where $Z \cdot e$ is the charge of the particle. This condition known as ``Hillas limit''~\citep{Hillas:1984ijl} can allow $E_{max} \gtrsim 10^{20}$ eV with $Z \sim 1$ for astrophysical environment like the ones present in neutron stars ($B \sim 10^{13}$ G, $L \sim 10$ Km)~\citep{Waxman:1997ti}, active galactic nuclei (AGNs) ($B \sim 10^{4}$ G, $L \sim 10$ AU)~\citep{Urry:1995mg}, radio lobes of AGNs ($B \sim 0.1 \mu$ G, $L \sim 10$ Kpc), and clusters of galaxies ($B \sim \mu$ G, $L \sim 100$ Kpc)~\citep{Berezinsky:1996wx}. The recent detection of a high-energy neutrino event ($\sim 290$ TeV) in coincidence with a gamma-ray flare, at hundred of GeVs, from the blazar TXS0506+056~\citep{IceCube:2018dnn,IceCube:2018cha} (at a distance of $z\sim 0.3365$~\citep{Paiano:2018qeq}) makes this class of sources one of the main candidates\citep{Murase:2014foa} of the diffuse astrophysical flux measured by IceCube~\citep{Aartsen:2016oji}. Even though this class of sources is extremely variable~\citep{2021MNRAS.tmp.1320M} the part of neutrinos, above 100 TeV, emitted during a flaring period can potentially be observed~\citep{Oikonomou:2019djc}. For blazars (like the sample reported in Fig.~\ref{fig:sensy1}) we can expect most of the UHE neutrinos produced through photohadronic interaction of accelerated protons with photons emitted by electrons within the jet. For other classes of AGN, like radio galaxies (as the sample reported in Fig.~\ref{fig:sensy2}), we can expect an additional UHE neutrino production inside the lobes, present at the end of the hundred Kpc jets, through proton-proton inelastic collisions~\citep{Fraija:2017jok}. A recent analysis taking into account the position of UHE cosmic-rays observed by Pierre Auger~\citep{ThePierreAuger:2015rma} and Telescope Array~\citep{AbuZayyad:2012ru} experiments, a known blazar gamma-ray catalog and the position of IceCube astrophysical events shows hints of possible correlation between the two messengers and highlights the fact that these class of accelerators have the capability of reaching the energy of $\sim 10^{20}$ eV. On this regards also clusters of galaxies can have the possibility to produce UHE neutrinos since cosmic rays accelerated up to UHE by AGN activities can be contained for a certain amount of time inside the cluster due to the presence of high magnetic field. They can reach values ranging from a few microgauss ($\mu$G) on scales of order $\sim 10$ Kpc for normal clusters~\citep{Clarke:2004te}, up to $10-40$ $\mu$G on scales of $3-5$ Kpc for cool core clusters~\citep{Ensslin:2005uk}. Such strong magnetic fields can confine cosmic-rays of UHE. While propagating inside the cluster, cosmic-rays can thus have interactions with the present photonic and baryonic backgrounds producing gamma-rays and neutrinos. \\Another possible candidate to produce UHE neutrinos it is represented by the newly born millisecond pulsars with a diffuse component reported in fig.~\ref{ANDIAMO_sensi}. The amount of this neutrino production can be traced though the measurements of the extra-galactic star formation rate following the model reported by~\citep{Fang:2013vla}.

\subsection{Extreme Dark Matter candidates}
The standard cosmological model is based on the existence of Dark Matter (DM) which, until now, did not show direct evidences via non-gravitational interaction. An alternative way to probe DM interactions, beyond the gravitational one, is the search for indirect signatures of DM decay or annihilation. These processes can indeed lead to the production of an astrophysical signal of cosmic-rays, gamma-rays and neutrinos.
In the last decades, the rise of high-energy multi-messenger astronomy has led to huge improvements in the indirect searches for dark matter.
Dark matter candidates concentrated in the galactic halo and distributed in the intergalactic space can produce a flux of UHE neutrinos through their decay and annihilation into ordinary matter. Due to the unitarity bound on DM cross-section~\citep{Griest:1989wd,Smirnov:2019ngs} we can expect a higher flux for the decaying case with respect to the annihilating one. The neutrino flux from decaying dark matter can be expressed by:
\begin{equation}
\phi_{\nu}^{dec}\sim \frac{\rho_{DM}}{m_{DM}}\frac{1}{\tau_{DM}},    
\end{equation}
where $\rho_{DM} \simeq 0.4 GeV/cm^{3}$ is the typical galactic density of DM particles, $\tau_{DM}$ is their lifetime, $m_{DM}$ is the DM mass, and $L\sim 10 kpc$ is the length scale of our galaxy.  On the other hand, for annihilating dark matter the corresponding flux is of the order of:
\begin{equation}
\phi_{\nu}^{ann}\sim \left(\frac{\rho_{DM}}{m_{DM}}\right)^{2} \sigma v_{DM} L,   
\end{equation}
where $\sigma$ is the annihilation cross-section, $v_{DM}\simeq 10^{-3} c$ corresponds to the typical velocity of dark matter particles and $L$ is the length scale. 
The possible values of neutrino fluxes obtained for annihilating DM result far beyond the sensitivity of next-generation neutrino radio telescopes, producing a number of neutrino events in these observatories negligibly small. An exception can be represented by peculiar scenarios featuring very dense and/or very cold dark matter substructures~\citep{Zavala:2014dla}. Conversely, for decaying DM particles with $\tau_{DM} \simeq 10^{29}$ s and $m_{DM} = 10^{9}$ GeV we can expect a production of neutrino flux within the reach of upcoming neutrino UHE telescopes~\citep{Chianese:2021htv}.

\section{Experimental setup}
\begin{figure}[h!]
\centering
\includegraphics[width=\textwidth]{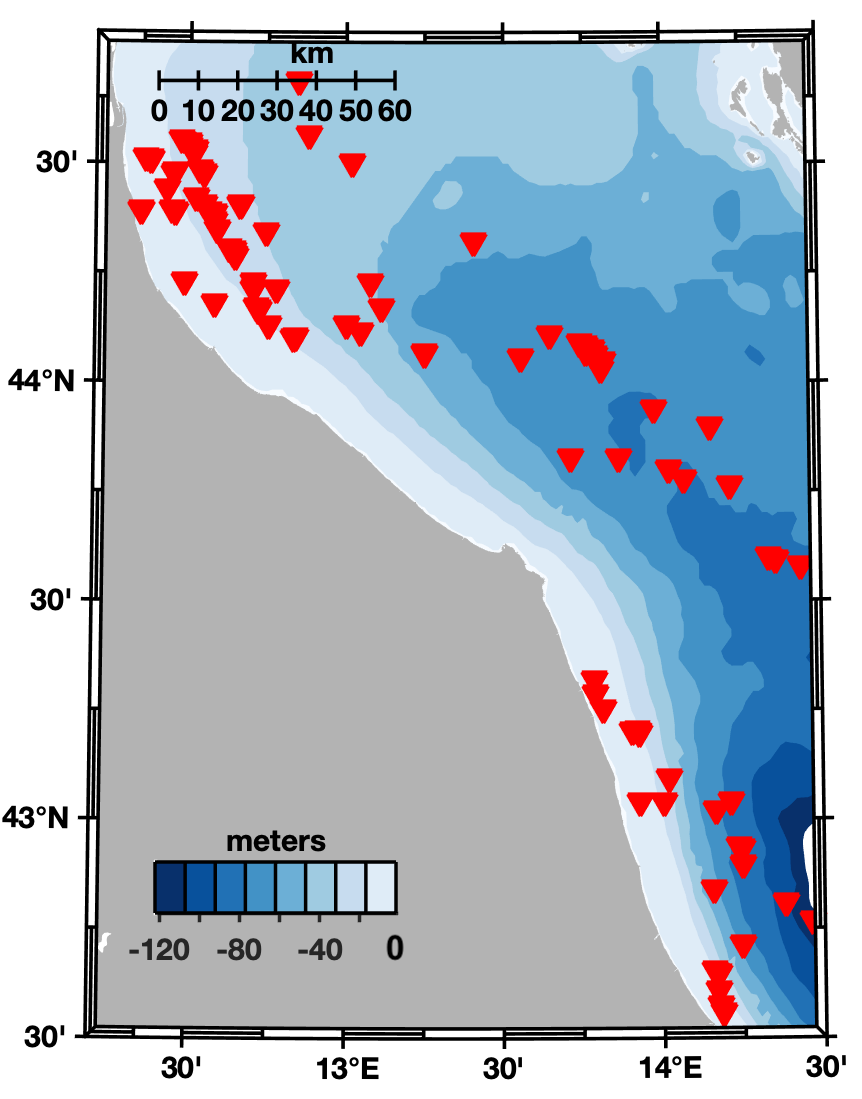}
\caption{In the map the red triangles show the location of the available platforms and colour coded the bathymetry of the Adriatic Sea.} 
\label{mappapiattaforme}
\end{figure}

The design of acoustic neutrinos detectors has been up to the present ancillary to the optical ones. Therefore, the larger attenuation length of the sound in water in respect to the light was never exploited properly. 
Here we explore the possibility of detecting a neutrino generated acoustic signal by instrumenting the area covered by the existing oil rigs structures.\\
As a best case scenario we consider all the available platforms (see Fig.~\ref{mappapiattaforme}) and based on the actual geometry we calculated the possible sensitivity of the acoustic telescope.
The study has been performed by accounting for sound attenuation in the conditions of Adriatic Sea and wave field propagation characteristics in the velocity structure scenario reported by \cite{misurevelocita}.
A ray tracing simulation is obtained using a realistic velocity profile for both the calculations of amplitude attenuation and ray propagation.  The dependence of wavefield propagation respect to the angle of the incident neutrino respect to the normal vector to the sea surface is studied here for the first time in shallow waters.
In exploration geophysics and in earthquake seismology, the use of large arrays of sensors is routine since decades. In both fields, the target is to study the structure of the soil to reveal potential geological structures that can trap oil or gas.
The detection of an acoustic signal with an extended array of sensors (hydrophones) is a problem that is solved since decades both in submarine sonar technology and in a similar fashion by array techniques for seismic studies and earthquake location like f-k analysis and beam-forming that are explained in e.g. \citep{https://doi.org/10.2312/gfz.nmsop-2ch9}. The problem in our case is made easier since the source is  well modelled by previous studies~\citep{Niess2006,BEVAN2009398}.\\
We plan thus to instrument 100 platforms, the ones reported with a red triangle in Fig.~\ref{mappapiattaforme}.
Each platform will be equipped with a multi-storey structure of hydrophones whose distribution in depth will be optimized according to the simulation of the wave field propagation studies. These will be based on the experimental data collected at the first R\&D stage.
In all platforms, a sub-array named single platform sub array (hereinafter SIPSA) will be deployed to better solve the f-k analysis uncertainties connected to spatial aliasing due to possible multiple reflection from the seafloor and water surface. 
We therefore plan to install a single projector in any array element (platform) to calibrate the SIPSA sub-arrays and the entire large detector. A continuous calibration using known signals resembling the neutrino expected one according to simulations see Fig.~\ref{wavef} will permit to develop many detection templates that will help to discriminate the signal and avoid false detections. In total, we will instrument 100 platforms (see Fig.~\ref{mappapiattaforme}) each of them will be instrumented with the SIPSA unit that can be formed by four to ten hydrophones.
A large advantage of shallow sea depth is indeed the simplicity of the hardware needed. In fact, a single hydrophone with an integrated pre-amplifier like~\citep{idrofono} has a self-noise level well below sea state zero and the integrated preamplifier permits the use of a simple analog cable. A cable no longer than 50 meters permits to use the analog output of the transducer and digitize it on board of the platform, avoiding the installation of an expensive underwater digitizer and the development of custom DAQ hardware. The reduction of the costs allows to multiply the number of sensors, increase the signal-to-noise ratio as well as the array sensitivity. 

\subsection{The existing infrastructure in Adriatic Sea} 
The platforms usable for the ANDIAMO experiment are located in the offshore of Adriatic Sea at a distance from the shore ranging from 10 to 60 kilometres. Most of them are placed in an elongated shape from north to south, following roughly the natural shape of the continental margin where oil and gas were extracted. The depth of the sea is  typical of a continental shelf, with depth ranging from 25 to 80 meters. 

The shallow depth of Adriatic Sea gives us several advantages, the most important is in terms of geometrical spreading. The acoustic wave generated by neutrino interaction never goes in the far field regime where spherical spreading is characteristic, this is the case for acoustic detectors implemented on optical arrays for exampele, where the array dimensions are small compared to the extension of the channel where they are installed. In our case, the cascade length is comparable to the depth of the channel, there is no transition region from near to far field propagation. The amplitude of the acoustic wave will decay as $1/\sqrt r$ respect to a spherical spreading where the decay is $1/r$. In terms of attenuation, we have 10 dB/decade against 20 dB/decade for spherical spreading.
In other words, we deal with a cylindrical wave instead of a spherical one.
The relatively high temperature of the shallow Adriatic Sea together with a lower salinity due to the fresh water influx from the largest Italian river, the Po river, will contribute to decrease the attenuation~\citep{libroneacustica}.
The Mediterranean Sea is for many physical and geological reasons the best environment for an underwater neutrino telescope as already highlighted by~\citep{Niess2006}.
Another advantage is logistical since mostly all services needed such power and connectivity are supposed to be already available on the oil rigs and no complex high pressure waterproof equipment is needed. 

The steady structure of the platform will permit to install the SIPSA mini array in every platform in a solution similar to the one adopted in~\citep{Aynutdinov:2012zz}.
From another side, we need to carefully study the underwater noise that could be a limiting factor. The stationary noise floor for our system is mostly dependent on sea state in the frequency range of detection for the predicted acoustic signal (1-20 kHz).
Transient noises as dolphins and propellers can of course spoil the sensitivity locally and in an unpredictable manner but in a long term perspective can be filtered and recognized since they have a different time /frequency signature compared to the neutrino generate ones.
Another possible issue with the shallow sea can be a complex ray path propagation and possible reflections and absorption, especially due to the interaction with sandy or clay seafloor that we will face in a specific paragraph of this paper.
Sound propagation in very shallow waters has been studied largely in literature by~\citep{Kuperman2004,SousaCosta2013,Chen2016,Porter2010} converging to stating that shallow waters propagation can be seen as a small scale problem respect to deep waters, for this reason we will apply ray tracing to simulate our environment.
Refraction and reflections are the physical phenomena that guide the direction of the wave field across the path from the source to the various receivers. The refraction is connected to the change of the vertical velocity profile that drives an acoustic impedance change. This profile changes across the year, and it depends on water temperature, salinity pH, and depth (pressure). For these reasons, a preliminary study using a realistic and appropriate velocity profile is necessary.
The structure of the velocity profile is indeed a game changer in the possibility to design an effective acoustic telescope, in particular we want to exploit the presence of the SOFAR channel to stretch the acoustic detection range to the maximum.
In fact, changing the physical properties of the medium changes the sensitivity of the detector itself. 
We will show that the available geometry of platforms distribution is suitable for the realization of the largest acoustic array for UHE neutrinos detection with a sensitive area of more than 10000 Km$^{2}$

\subsection{A novel reconstruction technique for acoustic signal} 
When a UHE neutrino interacts with matter at such high energies, it can decay according to different channels and generate an hadronic cascade that in turn gives rise to the thermo-acoustic effect.
\begin{figure}[h!]
   
     \begin{subfigure}[b]{0.53\textwidth}
         \centering
         \includegraphics[width=\textwidth]{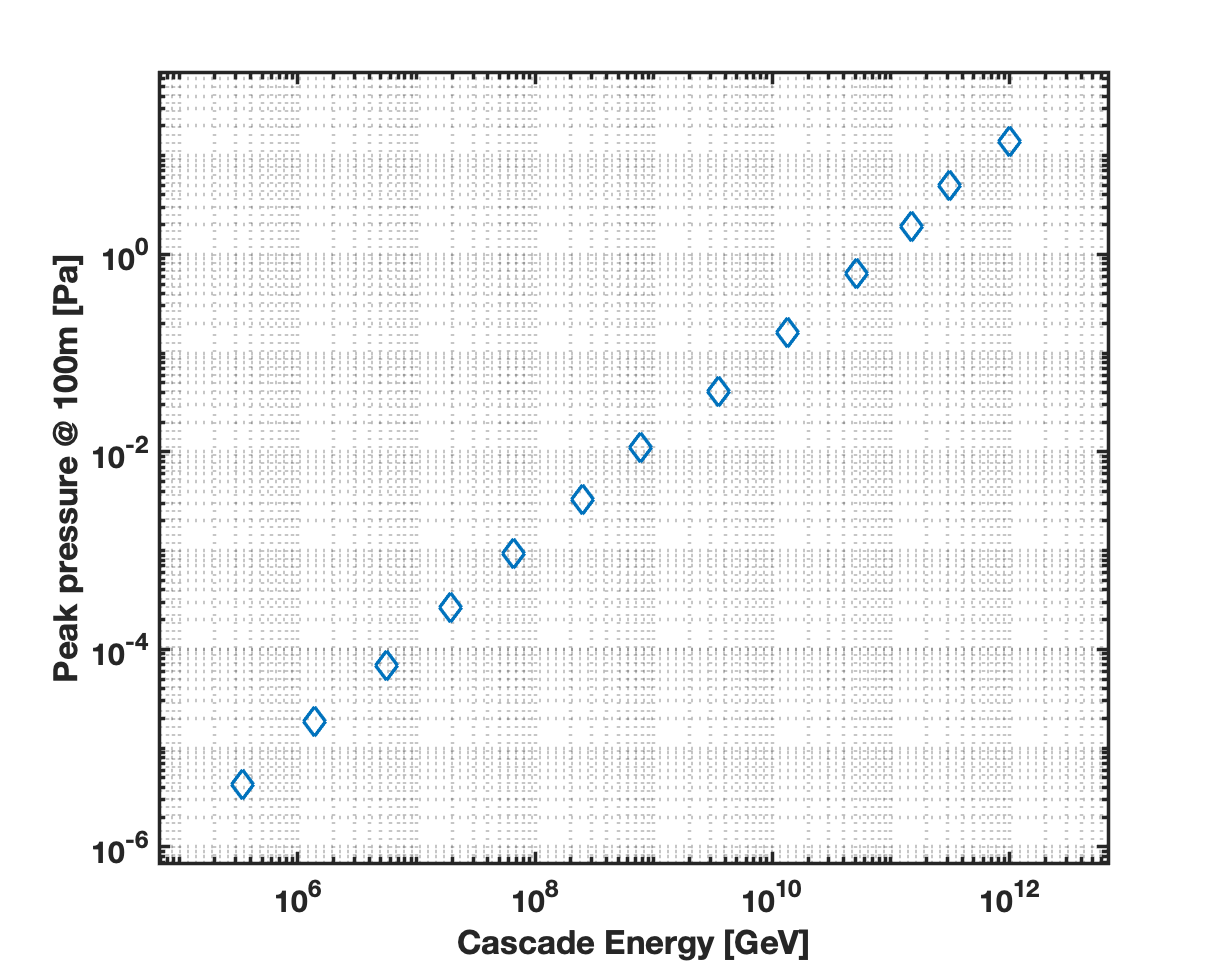}
         \caption{Peak pressure versus cascade energy calculated at 100 meters from the shower.}
         \label{pressdist}
     \end{subfigure}
     \hfill
     \begin{subfigure}[b]{0.53\textwidth}
         \centering
         \includegraphics[width=\textwidth]{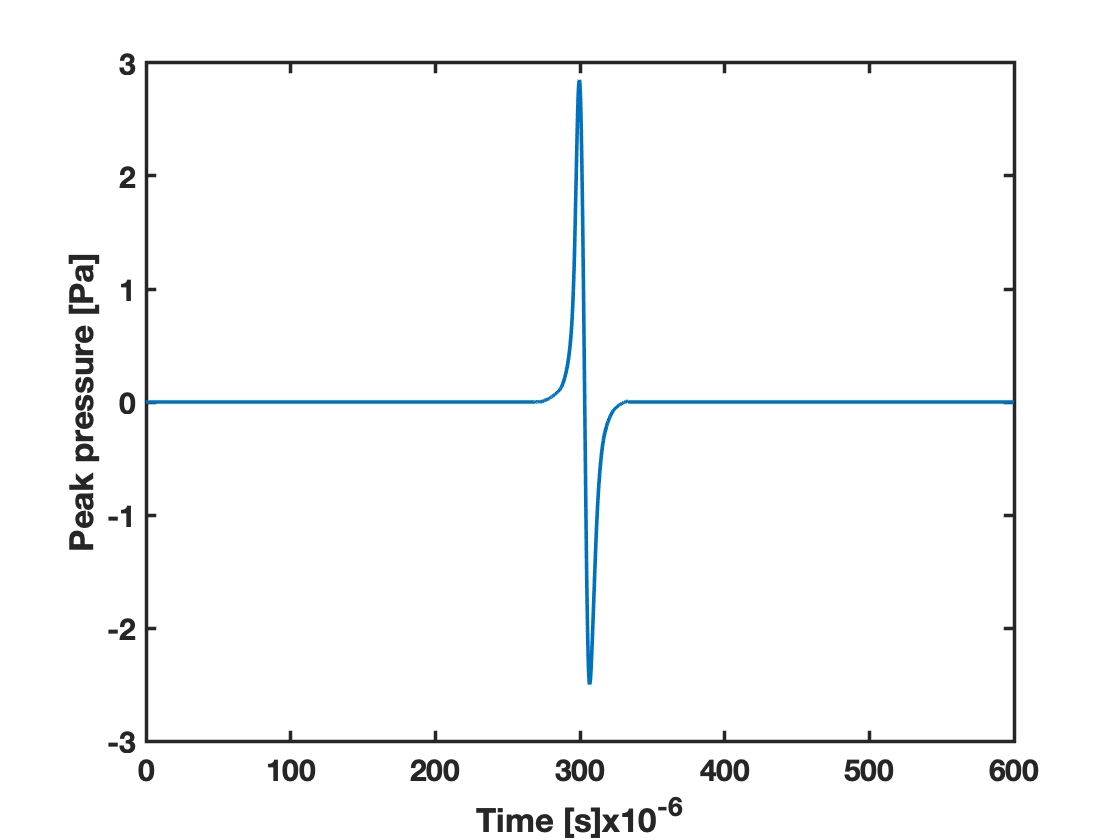}
         \caption{The waveform of the acoustic pulse at 100 m for a 1e11 GeV shower.}
         \label{wavef}
     \end{subfigure}
     \caption{Acoustic parameters of the source.}
    \end{figure}

The extension in depth of the cascade gives the length of the cylinder where the thermo-acoustic effect occurs. The intensity of this cylindrical wave, also called in literature pancake, depends on the angle of the cascade with respect to the vertical and of course on the generating phenomena, i.e. the possible decay channels. Once the acoustic wave is generated with a given intensity, the problem is shifted to a proper reconstruction and validation of the signal using the state-of-art acoustic techniques. The waveform generated by a UHE neutrino interaction resembles a bipolar signal, whose amplitude at 100 meters from the cascade can range from 10 Pascals to some milliPascal. The central frequency is expected to be around 50 kHz at 100 meters with a duration of the order of several microseconds according to simulations proposed by~\citep{BEVAN2009398}.
This energy deposition is similar to a delta function, exciting thus a white kind of spectrum in the frequency domain.
When this pulse propagates, it undergoes attenuation due to both geometrical spreading and absorption. This causes a broadening of the wavelet in time and an obvious decrease in amplitude.
Nowadays, state-of-the-art hydrophones have a self-noise level that is lower than the sea state level zero of the Wenz sea state plots \citep{wenz}. This means that, supposed to have the most sensitive hydrophones, we are left with the problem of a realistic calculation of the attenuation and frequency content change in the signal versus distance.  
This study is fundamental to assess how far can we stretch the sensitive area of the detector. 
The sea noise is a limiting factor to the sensitivity of the array. This will be discussed in detail in the section about the amplitude spectra. We will account for geometrical spreading and attenuation in the frequency domain. The generated signal creates in the far field (> 100 m) a cylindrical wave that propagates and attenuates with an $1/\sqrt r$ attenuation law. The frequency content of the bipolar signal ranges from 1kHz to 100 kHz and the peak pressure is expected at a frequency of approximately 50 kHz at 100 meters distance.
The signal undergoes all the acoustic phenomena involved in wave propagation in a vertically inhomogeneous medium, including refraction and dispersion/attenuation. The propagating pressure wave field changes the energy and energy distribution in the frequency domain with respect to distance from the event that generated it. It is no mystery that attenuation is a frequency dependent phenomenon and roughly speaking can be summarized with this statement: the longer wavelengths travels longer from the source. The previous considerations were made in order to answer to the basic question of this paper, i.e.: Can an array created by using the existing oil rigs (or at least a group of them) be suitable for the detection of UHE neutrinos ?
The results showing the feasibility of the ANDIAMO project can be grouped in two parts, the first regarding the attenuation/dispersion and the second one about the ray paths simulations addressing the following points:
\begin{itemize}
    \item Establish how far can we ear an acoustic signal generated by a neutrino induced shower in a relaistic velocity profile scenario.
    \item Determine which are the angular limits for detection in our setup. 
\end{itemize}

\subsubsection{Amplitude spectra}
\begin{figure}[h!]

\includegraphics[width=\textwidth]{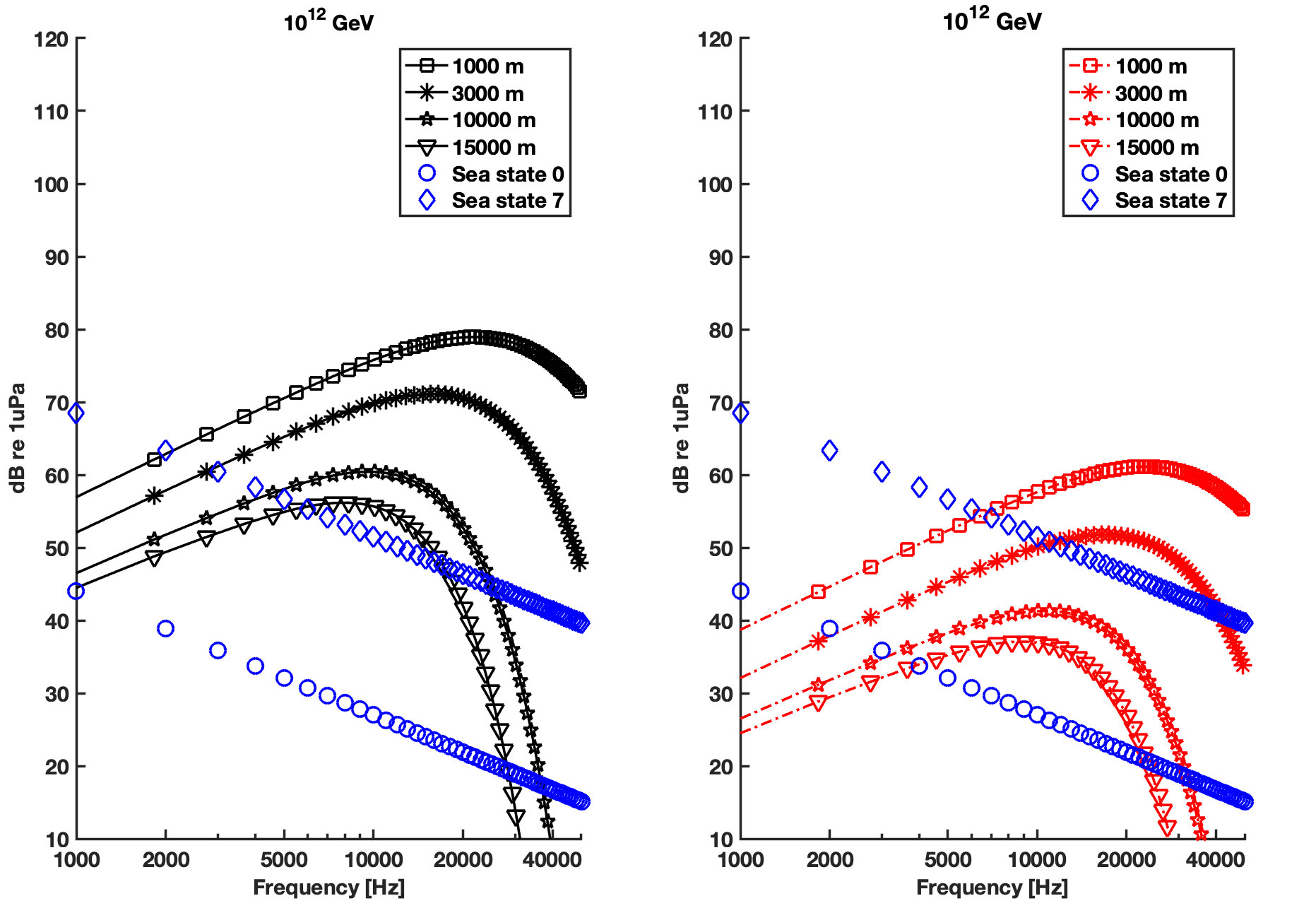}
\centering
\caption{Calculated spectra for different sea depths with fixed shower energy, on the left panel a 30 meters deep sea and on the right panel a 3000 meters deep sea, we assume isovelocity conditions.}
\label{cfrdepth}
\end{figure}

We calculate the amplitude spectrum of the acoustic pressure reference signal of Fig.~\ref{wavef} i.e. the one intended as the near source signal. 
\begin{figure}[h!]
\centering
\includegraphics[width=\textwidth]{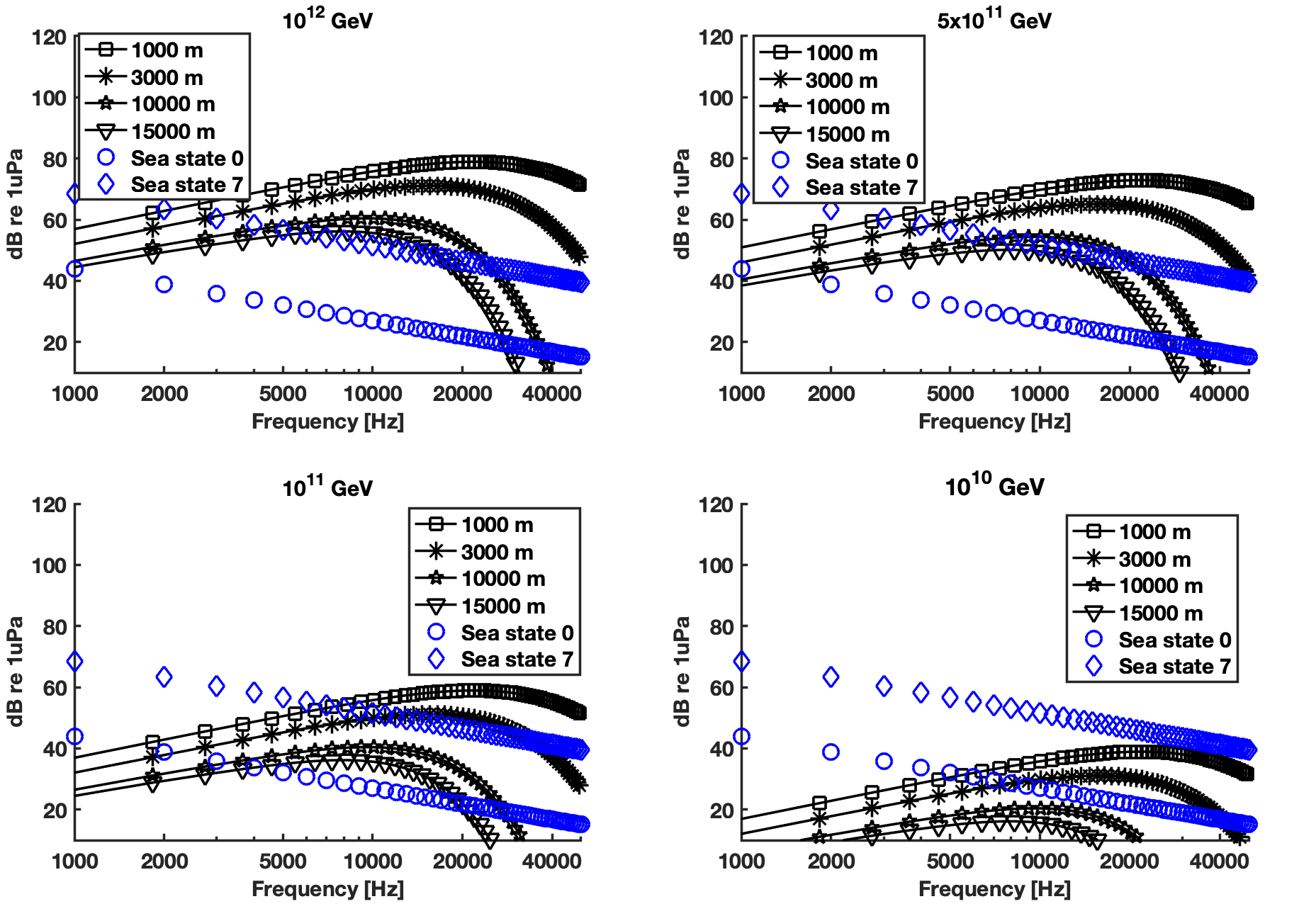}
\caption{Sound pressure level versus frequency and distance from the source for different shower energies in a 50 meters deep sea assuming isovelocity conditions.}
\label{rangespectrum}
\end{figure}
In a second step we calculate, for a fixed channel depth and for several distances and frequencies, the attenuated spectra for several source levels. The result is reported in Fig.~\ref{rangespectrum}.
As a comparison, for a fixed energy, we calculate the dispersion of the spectrum for a 3000 meters deep channel to highlight the advantage of shallow waters (see Fig.~\ref{cfrdepth}).
From the spectra, it is evident that above a cascade energy of $10^{19}$ GeV signal can be measured by any single array element (oil rig) that stays in a 10-kilometre radius from the source event.
We assume a linear path propagation from source to receiver induced by an isovelocity condition for simplicity. 
We calculate the transmission losses (TL) for every spectral component of the source signal reported in Fig.~\ref{wavef} and using the following equation:
\begin{equation}
    TL=10 \log \frac{I_s}{I(r)}
    \label{tloss}
\end{equation}

we obtain in the end the sound pressure level spectra at different distances for different source energies and incidence angle of 0 deg, the angular dependence of attenuation will be studied in the second part where we will account for seafloor reflection.
We use as source level the sound pressure level at 100 meters because it can be considered far field compared to the extension of the neutrino induced cascade (that is not a point-like source) but still much smaller than the inter distance occurring between the array elements.
An extrapolation of the results shown by~\cite{BEVAN2009398} permitted to produce a plot (see Fig.~\ref{pressdist}) of peak-pressure versus cascade energy at a fixed distance of 100 m which is more suitable for our study.
The results of the simulated spectra for a 50 meters deep channel for several energies and distances from the source are reported in Fig.~\ref{rangespectrum}.

\subsubsection{Ray tracing and signal intensity versus cascade angle}

The problem of propagation in shallow waters, i.e. inside the continental shelf, can be complex, and it depends strongly on the velocity structure of the water column. 
\begin{figure}[h!]
\centering
\includegraphics[width=1.05\textwidth]{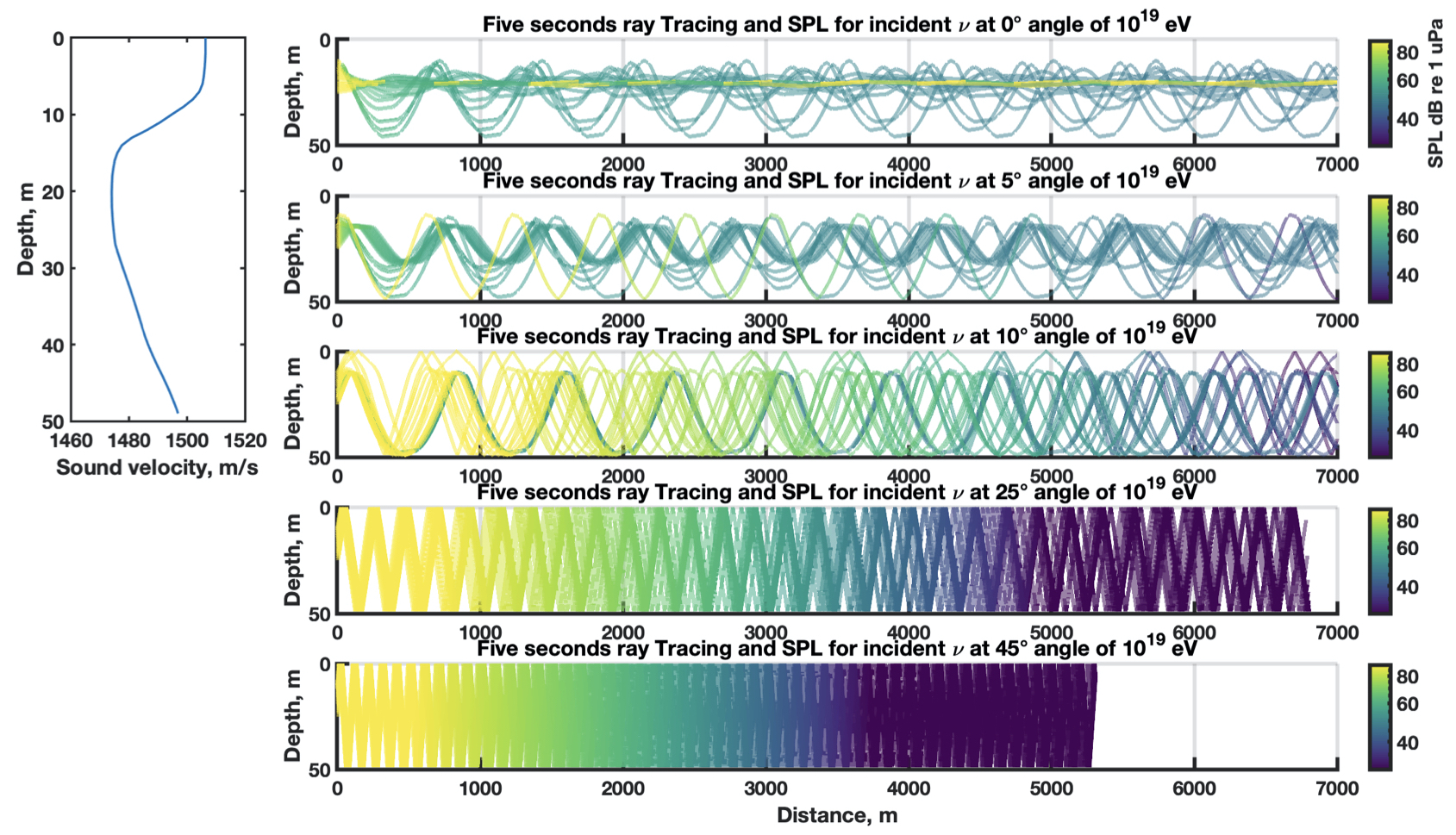}
\caption{Ray tracing using a typical velocity profile for a 50 m deep sea at different neutrino incidence angle and a shower energy of $10^{19}$ eV.}
\label{raytangles}
\end{figure}
In our setup, the seafloor is pretty homogeneous, and we have a general mild increase of the seafloor depth from 20 to 50 meters according to the natural slope of the continental shelf.
Future studies based on the geophysical and oceanographic acquired data of the detector site will permit to produce more accurate simulations of the propagation of the signals. 
\begin{figure}[h!]
    \centering
    \includegraphics[width=1.05\textwidth]{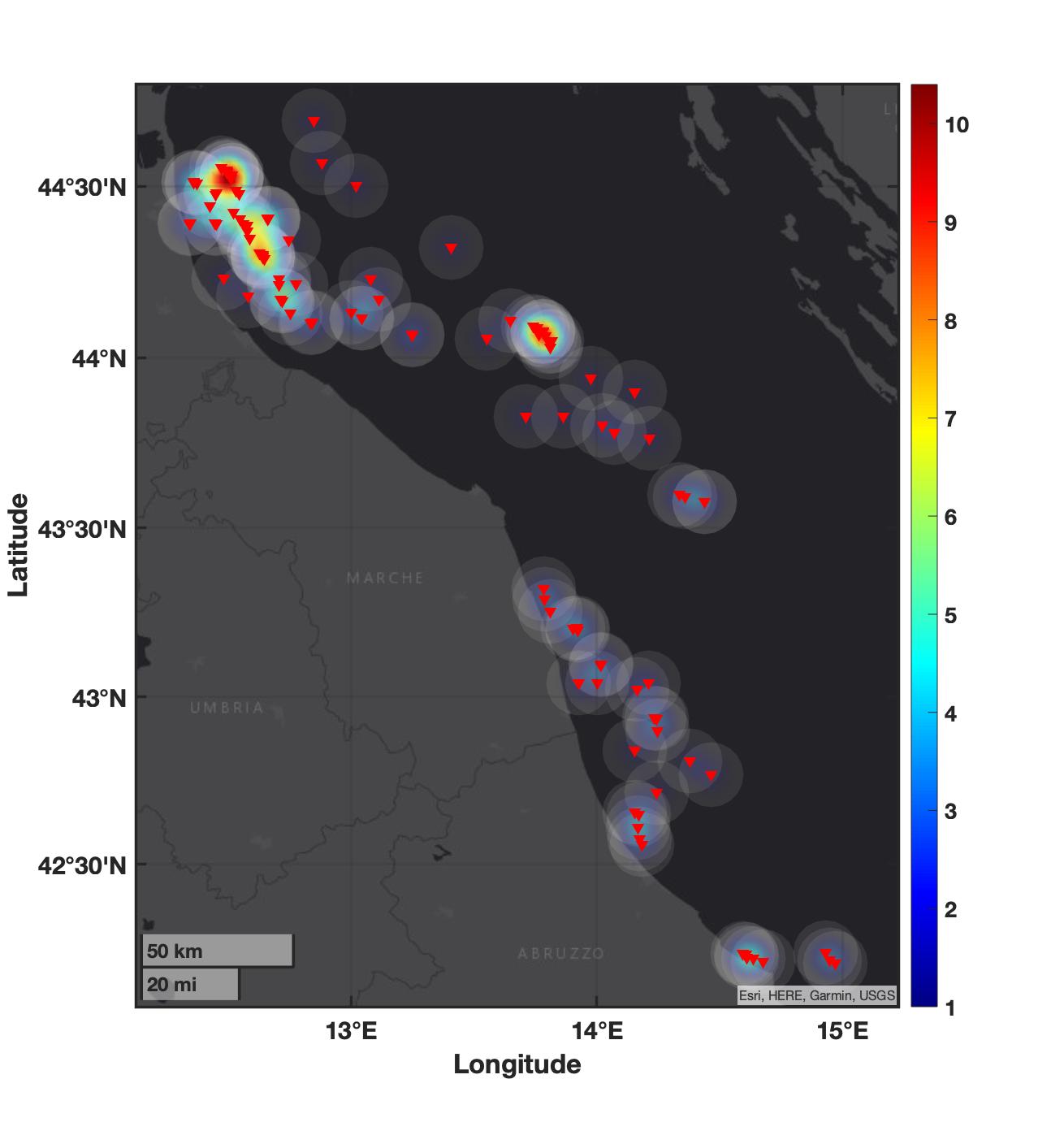}
    \caption{Map illustrating the covered sensitive area for vertical neutrino induced showers at $10^{19}$ eV , color coded is the SIPSA sub array receiver density}
    \label{zerodegmap}
\end{figure}
The typical shallow water conditions are studied largely in underwater acoustics~\cite{Kuperman2004,SousaCosta2013} we can summarize the part of our interest with the following scenario that is confirmed by experimental data e.g. \cite{misurevelocita} for Mediterranean sea. In shallow waters we have a first mixed surface layer where the speed of sound is constant, followed by a thermocline and an increase due to pressure as shown in the left top panel of Fig.~\ref{raytangles}. During winter the entire water column is completely mixed, and the sun exposure is lower. We have a less steep thermal gradient and a larger mixing due to the heavy sea conditions. This creates a condition that can be more similar to isovelocity conditions. The latter implies the straight ray propagation given the homogeneity of the medium. By simulating a cylindrical source of 20 meters of extension in depth (cascade length), we could ray-trace the direction of the wave-fronts generated at different shower angle for a 5 kHz tone which is, according to spectral simulations, the peak level at the distances of interest given the actual platforms geometry. Additional signal attenuation induced by transmission losses at the sea floor are accounted for rays that are not confined inside the SOFAR channel. The transmission losses at the sea floor constitute a severe limiting factor for angles larger than 45 degrees since after this value we have a very steep drop in the reflection coefficient. We consider the sea surface as a perfect reflector, hence we account only for sea bottom losses. We report the results in Fig.~\ref{raytangles}, we can easily note that at after 3000 meters for a shower induced at 45$^{\circ}$ the signal drops below the sea state level zero value for a 5kHz tone. On the contrary for angles between $\pm 10^{\circ}$ the channeling effect permits the focusing of the energy inside the the SOFAR channel and extending the range of the detector. This aspect of wave propagation is limiting the effective sky solid angle that we can observe with the current setup. Of course a denser array can push further the angular span that we can cover.
In summary the attenuation, geometrical spreading and reflection/transmission losses are carefully calculated in the simulation of the acoustic environment of the ANDIAMO experiment. From the acoustical point of view the sensitive area changes according to angle and energy of the incident neutrino an example of the covered area is reported in Fig. \ref{zerodegmap} where the color map represent the SIPSA sub arrays density. In this case for a $10^{19}$ eV cascade vertically originated the sensitive area is of the order of 10000 km$^2$.

\section{Possible sensitivity to UHE neutrinos}

\begin{figure}[h!]
\centering
\includegraphics[scale=0.80]{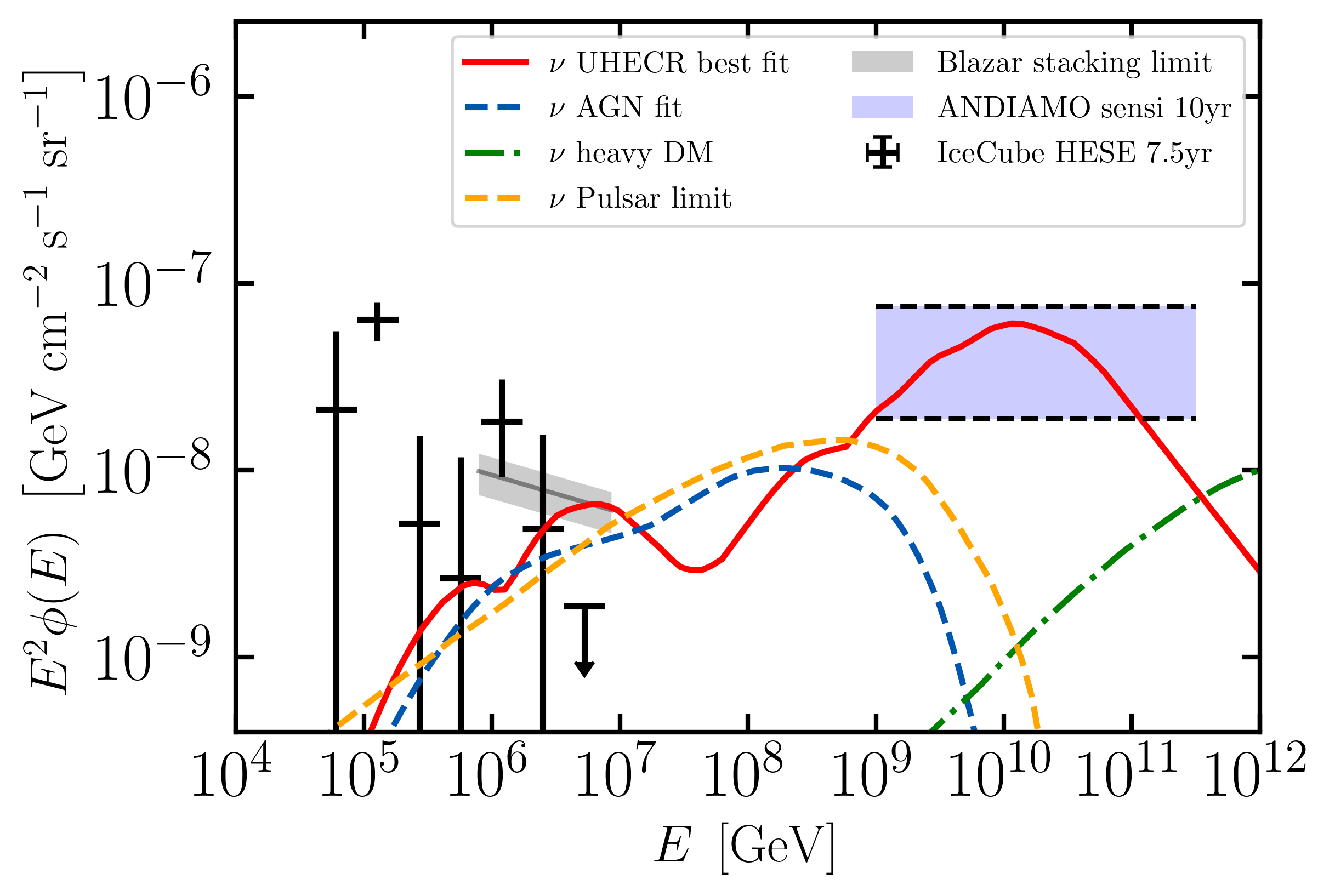}
\caption{In this plot we report the expected sensitivity of the ANDIAMO concept with a surface array equivalent to the one described in Fig.~\ref{mappapiattaforme} and a observational time of one decade. The three neutrino families are considered with the range of sensitivity spanning the different sensitive areas covered considering the possible beneficial incident angles. For comparison we show also the limit on the possible UHE cosmogenic neutrino flux obtained from the UHECR observations as well as possible UHE neutrino diffuse contributions from AGNs, millisecond pulsars and heavy dark matter decays. IceCube full-sky measurements are reported too.}
\label{ANDIAMO_sensi}
\end{figure}

Since the calculation of a detailed exposure for the proposed ANDIAMO array go beyond the scope of this work, mainly because of a non-defined geometry and the absence of a effective area obtained though a dedicated Monte Carlo simulation, a possible approximation is obtained through a semi-analytic analysis.
The limits for the angle and area are set by the results shown in the previous section where acoustic propagation is studied in detail.
The maximal solid angle around the zenith of the detector is set by the abrupt fall of the reflection coefficient which in turn limits the total distance travelled by the acoustic signal generated by an inclined cascade.
The sensitive area is then angle and energy dependent, for a $10^{19}$ eV cascade it ranges from 10000 Km$^2$ for a vertical cascade to about 2800 Km$^2$ for a 45$^{\circ}$ inclined one.
This area variability is accounted in the calculation of the sensitivity shown in Fig.~\ref{ANDIAMO_sensi}.

For downward-going neutrinos, the calculation of the exposure involves the ANDIAMO array aperture, the neutrino interaction probability, an ideal identification efficiency, and the integration in time.
A sum over time and integration in solid angle would yield the exposure ($\mathcal{E}$) to UHE neutrinos.
Assuming a $1 : 1 : 1$ flavour ratio (as expected due to the effects of neutrino oscillations during propagation from the sources to the earth), the total exposure can be expressed as:
\begin{equation}
\label{exposure}
\mathcal{E}(E_{\nu}) = \frac{2\pi}{m}\sum_{i}\left[\sigma_{i}(E_{\nu}) \int dt d\theta dD \sin \theta \cos \theta A^{i}_{eff}(\theta,D,E_{nu},t)\right] ,
\end{equation}
where the sum runs over the three neutrino flavours and the CC and NC interactions, with $\sigma^{i}$ the corresponding $\nu-$nucleon interaction cross-section~\citep{Cooper-Sarkar:2007zsa} and $m$ the nucleon mass. The integral is performed over the zenith angle $\theta$ the
interaction depth $D$ of the neutrino (in units of g cm$^{-2}$), and
the blind search period. $A^{i}_{eff}$ represents the effective area. Without having a finalized geometry for the ANDIAMO array, we did a analytic approach to obtain a possible effective area and a related sensitivity of this future detector. In particular, we assume $A^{i}_{eff}$ to be approximated by $Ap_{\nu \rightarrow \mu,e,\tau}(E_{\nu},\langle E_{\mu,e,\tau}\rangle$) for an array of surface $A$ (from 2800 to 10000 Km$^{2}$ for the ANDIAMO case). Where $p_{\nu \rightarrow \mu,e,\tau}(E_{\nu})$ can be described by the following equation:
\begin{equation}
\label{effarea}
p_{\nu \rightarrow \mu,e,\tau}(E_{\nu}) = R_{\mu,e,\tau}(\langle E_{\mu, e, \tau}\rangle)/\lambda_{\nu}(E_{\nu}),
\end{equation}
with $\lambda_{\nu}(E_{\nu})$ representing the interaction length of a neutrino with energy $E_{\nu}$ and $R_{\mu,e,\tau}$ describing the leptonic range in the considered medium. These quantities allow us to roughly obtain a possible flux limit observable, or in other words, a possible sensitivity of our detector. More in details, under the assumption that the UHE neutrino flux behaves like $\Phi(E)=\frac{dN}{dE}=kE^{-2}$ the corresponding sensitivity can be written as:
\begin{equation}
\label{sensi}
k = \frac{N_{evt}}{\int^{E_{max}}_{E_{min}}E^{-2}\mathcal{E}(E)dE},
\end{equation}
where the $\mathcal{E}(E)$ correspond to the exposure obtained in equation~\ref{exposure} and $N_{evt}$ indicates the number of expected events. No background events neither spectral noises were considered in the energy range where ANDIAMO array can have the capability of reconstructing UHE neutrinos.\\

The expected sensitivity, reported in Fig.~\ref{ANDIAMO_sensi}, is compared with the main expected UHE neutrino diffuse fluxes. Event though more accurate studies can be obtained with the use of a dedicated Monte Carlo simulation chain whenever the geometry will be finalized, the preliminary calculations obtained through an analytical approach highlights the potential discoveries of such acoustic array concept.


\section{Conclusions}
The advances of neutrino astronomy of the last decade surge the importance of knowing the neutrino spectral features in the energy range from PeV to EeV. Several radio arrays prototypes have been recently finalized with the idea of covering larger surfaces in order to be sensible to the possible extraterrestrial UHE neutrino flux. An alternative way to detect this flux it is represented by the underwater/ice acoustic techniques carried on in the last decades with small prototypes or through the possible use of hydrophones installed for calibration purposes and not optimized for an acoustic detector design. In this work we explore the possibility of building a dedicated acoustic array in the Adriatic Sea using the platforms already installed, and not operative, for oil and gas upstream.\\
Detailed calculations on the propagation and attenuation of the signal generated by UHE neutrinos in shallow waters show the possibility to exploit the favourable conditions of the Adriatic Sea. This advantage is  given by the physical and geographical conditions of the sea, i.e. mild temperature and low salinity. Also, the existence of many infrastructures partially instrumented is a plus. Indeed, it will permit the deployment of the largest acoustic array for UHE neutrinos detection as well as for marine biology, oceanographic and geophysical studies. The calculations performed show that it is feasible to realize an acoustic telescope with an instrumented maximal area of $\sim$10000 Km$^2$ exploiting all the available platforms all over the Adriatic Sea with an expected sensitivity shown in Fig.~\ref{ANDIAMO_sensi}. A first attempt to estimate the possible sensitivity shows that, even though the possibility of resolving single sources of UHE neutrinos is still a challenging purpose, constraining the neutrino spectral features in the PeV-EeV range can be possible within a decade of data taking.\\
Considering the moderate costs of this detection technique and the uniqueness of this project i.e. complementary to the shower earth-skimming radio array, we aim to make further studies with a dedicated Monte Carlo simulation optimizing the geometry of hydrophone lines and analysing more in details the possible backgrounds with in situ measurements. Moreover, we leave open the future possibility of taking into account a bigger instrumental area with additional lines equipped without pre-existing platforms.

\section*{Acknowledgements}
The authors would like to thank Giorgio Riccobene, Robert Lahmann, Kay Graf and Giorgio Gratta for their helpful suggestions
and discussions regarding the detector feasibility study.
This work was partially supported by the research grant number 2017W4HA7S “NAT-NET: Neutrino
and Astroparticle Theory Network” under the program PRIN 2017 funded by the Italian Ministero
dell’Universit`a e della Ricerca (MUR).
\clearpage
\bibliography{references}

\begin{thebibliography}{10}
\expandafter\ifx\csname url\endcsname\relax
  \def\url#1{\texttt{#1}}\fi
\expandafter\ifx\csname urlprefix\endcsname\relax\def\urlprefix{URL }\fi
\expandafter\ifx\csname href\endcsname\relax
  \def\href#1#2{#2} \def\path#1{#1}\fi

\bibitem{Abraham:2010mj}
J.~Abraham, et~al., {Measurement of the Energy Spectrum of Cosmic Rays above
  $10^{18}$ eV Using the Pierre Auger Observatory}, Phys. Lett. B 685 (2010)
  239--246.
\newblock \href {http://arxiv.org/abs/1002.1975} {\path{arXiv:1002.1975}},
  \href {http://dx.doi.org/10.1016/j.physletb.2010.02.013}
  {\path{doi:10.1016/j.physletb.2010.02.013}}.

\bibitem{AbuZayyad:2012ru}
T.~Abu-Zayyad, et~al., {The Cosmic Ray Energy Spectrum Observed with the
  Surface Detector of the Telescope Array Experiment}, Astrophys. J. Lett. 768
  (2013) L1.
\newblock \href {http://arxiv.org/abs/1205.5067} {\path{arXiv:1205.5067}},
  \href {http://dx.doi.org/10.1088/2041-8205/768/1/L1}
  {\path{doi:10.1088/2041-8205/768/1/L1}}.

\bibitem{Halzen:2002pg}
F.~Halzen, D.~Hooper, {High-energy neutrino astronomy: The Cosmic ray
  connection}, Rept. Prog. Phys. 65 (2002) 1025--1078.
\newblock \href {http://arxiv.org/abs/astro-ph/0204527}
  {\path{arXiv:astro-ph/0204527}}, \href
  {http://dx.doi.org/10.1088/0034-4885/65/7/201}
  {\path{doi:10.1088/0034-4885/65/7/201}}.

\bibitem{Bahcall:1999yr}
J.~N. Bahcall, E.~Waxman, {High-energy astrophysical neutrinos: The Upper bound
  is robust}, Phys. Rev. D 64 (2001) 023002.
\newblock \href {http://arxiv.org/abs/hep-ph/9902383}
  {\path{arXiv:hep-ph/9902383}}, \href
  {http://dx.doi.org/10.1103/PhysRevD.64.023002}
  {\path{doi:10.1103/PhysRevD.64.023002}}.

\bibitem{Kotera:2010yn}
K.~Kotera, D.~Allard, A.~Olinto, {Cosmogenic Neutrinos: parameter space and
  detectabilty from PeV to ZeV}, JCAP 10 (2010) 013.
\newblock \href {http://arxiv.org/abs/1009.1382} {\path{arXiv:1009.1382}},
  \href {http://dx.doi.org/10.1088/1475-7516/2010/10/013}
  {\path{doi:10.1088/1475-7516/2010/10/013}}.

\bibitem{Aartsen:2014gkd}
M.~Aartsen, et~al., {Observation of High-Energy Astrophysical Neutrinos in
  Three Years of IceCube Data}, Phys. Rev. Lett. 113 (2014) 101101.
\newblock \href {http://arxiv.org/abs/1405.5303} {\path{arXiv:1405.5303}},
  \href {http://dx.doi.org/10.1103/PhysRevLett.113.101101}
  {\path{doi:10.1103/PhysRevLett.113.101101}}.

\bibitem{Aartsen:2015rwa}
M.~Aartsen, et~al., {Evidence for Astrophysical Muon Neutrinos from the
  Northern Sky with IceCube}, Phys. Rev. Lett. 115~(8) (2015) 081102.
\newblock \href {http://arxiv.org/abs/1507.04005} {\path{arXiv:1507.04005}},
  \href {http://dx.doi.org/10.1103/PhysRevLett.115.081102}
  {\path{doi:10.1103/PhysRevLett.115.081102}}.

\bibitem{Aartsen:2013jdh}
M.~Aartsen, et~al., {Evidence for High-Energy Extraterrestrial Neutrinos at the
  IceCube Detector}, Science 342 (2013) 1242856.
\newblock \href {http://arxiv.org/abs/1311.5238} {\path{arXiv:1311.5238}},
  \href {http://dx.doi.org/10.1126/science.1242856}
  {\path{doi:10.1126/science.1242856}}.

\bibitem{Murase:2005hy}
K.~Murase, S.~Nagataki, {High energy neutrino emission and neutrino background
  from gamma-ray bursts in the internal shock model}, Phys. Rev. D 73 (2006)
  063002.
\newblock \href {http://arxiv.org/abs/astro-ph/0512275}
  {\path{arXiv:astro-ph/0512275}}, \href
  {http://dx.doi.org/10.1103/PhysRevD.73.063002}
  {\path{doi:10.1103/PhysRevD.73.063002}}.

\bibitem{Biehl:2017zlw}
D.~Biehl, D.~Boncioli, A.~Fedynitch, W.~Winter, {Cosmic-Ray and Neutrino
  Emission from Gamma-Ray Bursts with a Nuclear Cascade}, Astron. Astrophys.
  611 (2018) A101.
\newblock \href {http://arxiv.org/abs/1705.08909} {\path{arXiv:1705.08909}},
  \href {http://dx.doi.org/10.1051/0004-6361/201731337}
  {\path{doi:10.1051/0004-6361/201731337}}.

\bibitem{Beresinsky:1969qj}
V.~Berezinsky, G.~Zatsepin, {Cosmic rays at ultrahigh-energies (neutrino?)},
  Phys. Lett. B 28 (1969) 423--424.
\newblock \href {http://dx.doi.org/10.1016/0370-2693(69)90341-4}
  {\path{doi:10.1016/0370-2693(69)90341-4}}.

\bibitem{IceCube:2016zyt}
M.~G. Aartsen, et~al., {The IceCube Neutrino Observatory: Instrumentation and
  Online Systems}, JINST 12~(03) (2017) P03012.
\newblock \href {http://arxiv.org/abs/1612.05093} {\path{arXiv:1612.05093}},
  \href {http://dx.doi.org/10.1088/1748-0221/12/03/P03012}
  {\path{doi:10.1088/1748-0221/12/03/P03012}}.

\bibitem{ANTARES:2011hfw}
M.~Ageron, et~al., {ANTARES: the first undersea neutrino telescope}, Nucl.
  Instrum. Meth. A 656 (2011) 11--38.
\newblock \href {http://arxiv.org/abs/1104.1607} {\path{arXiv:1104.1607}},
  \href {http://dx.doi.org/10.1016/j.nima.2011.06.103}
  {\path{doi:10.1016/j.nima.2011.06.103}}.

\bibitem{Aynutdinov:2012zz}
V.~Aynutdinov, et~al., {Acoustic search for high-energy neutrinos in the Lake
  Baikal: Results and plans}, Nucl. Instrum. Meth. A 662 (2012) S210--S215.
\newblock \href {http://dx.doi.org/10.1016/j.nima.2010.11.153}
  {\path{doi:10.1016/j.nima.2010.11.153}}.

\bibitem{KM3Net:2016zxf}
S.~Adrian-Martinez, et~al., {Letter of intent for KM3NeT 2.0}, J. Phys. G
  43~(8) (2016) 084001.
\newblock \href {http://arxiv.org/abs/1601.07459} {\path{arXiv:1601.07459}},
  \href {http://dx.doi.org/10.1088/0954-3899/43/8/084001}
  {\path{doi:10.1088/0954-3899/43/8/084001}}.

\bibitem{Askaryan:1962hbi}
G.~Askar'yan, {Excess negative charge of an electron-photon shower and its
  coherent radio emission}, Sov. Phys. JETP 14~(2) (1962) 441--443.

\bibitem{Kravchenko:2001id}
I.~Kravchenko, et~al., {Performance and simulation of the RICE detector},
  Astropart. Phys. 19 (2003) 15--36.
\newblock \href {http://arxiv.org/abs/astro-ph/0112372}
  {\path{arXiv:astro-ph/0112372}}, \href
  {http://dx.doi.org/10.1016/S0927-6505(02)00194-9}
  {\path{doi:10.1016/S0927-6505(02)00194-9}}.

\bibitem{Kravchenko:2011im}
I.~Kravchenko, S.~Hussain, D.~Seckel, D.~Besson, E.~Fensholt, J.~Ralston,
  J.~Taylor, K.~Ratzlaff, R.~Young, {Updated Results from the RICE Experiment
  and Future Prospects for Ultra-High Energy Neutrino Detection at the South
  Pole}, Phys. Rev. D 85 (2012) 062004.
\newblock \href {http://arxiv.org/abs/1106.1164} {\path{arXiv:1106.1164}},
  \href {http://dx.doi.org/10.1103/PhysRevD.85.062004}
  {\path{doi:10.1103/PhysRevD.85.062004}}.

\bibitem{Allison:2015eky}
P.~Allison, et~al., {Performance of two Askaryan Radio Array stations and first
  results in the search for ultrahigh energy neutrinos}, Phys. Rev. D 93~(8)
  (2016) 082003.
\newblock \href {http://arxiv.org/abs/1507.08991} {\path{arXiv:1507.08991}},
  \href {http://dx.doi.org/10.1103/PhysRevD.93.082003}
  {\path{doi:10.1103/PhysRevD.93.082003}}.

\bibitem{Barwick:2014rca}
S.~Barwick, et~al., {Design and Performance of the ARIANNA HRA-3 Neutrino
  Detector Systems}, IEEE Trans. Nucl. Sci. 62~(5) (2015) 2202--2215.
\newblock \href {http://arxiv.org/abs/1410.7369} {\path{arXiv:1410.7369}},
  \href {http://dx.doi.org/10.1109/TNS.2015.2468182}
  {\path{doi:10.1109/TNS.2015.2468182}}.

\bibitem{Barwick:2014pca}
S.~Barwick, et~al., {A First Search for Cosmogenic Neutrinos with the ARIANNA
  Hexagonal Radio Array}, Astropart. Phys. 70 (2015) 12--26.
\newblock \href {http://arxiv.org/abs/1410.7352} {\path{arXiv:1410.7352}},
  \href {http://dx.doi.org/10.1016/j.astropartphys.2015.04.002}
  {\path{doi:10.1016/j.astropartphys.2015.04.002}}.

\bibitem{Aguilar:2020xnc}
J.~Aguilar, et~al., {Design and Sensitivity of the Radio Neutrino Observatory
  in Greenland (RNO-G)}\href {http://arxiv.org/abs/2010.12279}
  {\path{arXiv:2010.12279}}.

\bibitem{2020SCPMA..6319501A}
J.~{{\'A}lvarez-Mu{\~n}iz}, et~al., {The Giant Radio Array for Neutrino
  Detection (GRAND): Science and design}, Science China Physics, Mechanics, and
  Astronomy 63~(1) (2020) 219501.
\newblock \href {http://arxiv.org/abs/1810.09994} {\path{arXiv:1810.09994}},
  \href {http://dx.doi.org/10.1007/s11433-018-9385-7}
  {\path{doi:10.1007/s11433-018-9385-7}}.

\bibitem{Gorham:2008dv}
P.~Gorham, et~al., {The Antarctic Impulsive Transient Antenna Ultra-high Energy
  Neutrino Detector Design, Performance, and Sensitivity for 2006-2007 Balloon
  Flight}, Astropart. Phys. 32 (2009) 10--41.
\newblock \href {http://arxiv.org/abs/0812.1920} {\path{arXiv:0812.1920}},
  \href {http://dx.doi.org/10.1016/j.astropartphys.2009.05.003}
  {\path{doi:10.1016/j.astropartphys.2009.05.003}}.

\bibitem{Allison:2019xtn}
P.~Allison, et~al., {Constraints on the diffuse flux of ultrahigh energy
  neutrinos from four years of Askaryan Radio Array data in two stations},
  Phys. Rev. D 102~(4) (2020) 043021.
\newblock \href {http://arxiv.org/abs/1912.00987} {\path{arXiv:1912.00987}},
  \href {http://dx.doi.org/10.1103/PhysRevD.102.043021}
  {\path{doi:10.1103/PhysRevD.102.043021}}.

\bibitem{Kurahashi:2010ei}
N.~Kurahashi, J.~Vandenbroucke, G.~Gratta, {Search for Acoustic Signals from
  Ultra-High Energy Neutrinos in 1500 km$^3$ of Sea Water}, Phys. Rev. D 82
  (2010) 073006.
\newblock \href {http://arxiv.org/abs/1007.5517} {\path{arXiv:1007.5517}},
  \href {http://dx.doi.org/10.1103/PhysRevD.82.073006}
  {\path{doi:10.1103/PhysRevD.82.073006}}.

\bibitem{Lehtinen2002}
N.~G. Lehtinen, S.~Adam, G.~Gratta, T.~K. Berger, M.~J. Buckingham,
  \href{https://doi.org/10.1016/s0927-6505(01)00158-x}{Sensitivity of an
  underwater acoustic array to ultra-high energy neutrinos}, Astroparticle
  Physics 17~(3) (2002) 279--292.
\newblock \href {http://dx.doi.org/10.1016/s0927-6505(01)00158-x}
  {\path{doi:10.1016/s0927-6505(01)00158-x}}.
\newline\urlprefix\url{https://doi.org/10.1016/s0927-6505(01)00158-x}

\bibitem{Askarian:1979zs}
G.~Askarian, B.~Dolgoshein, A.~Kalinovsky, N.~Mokhov, {ACOUSTIC DETECTION OF
  HIGH-ENERGY PARTICLE SHOWERS IN WATER}, Nucl. Instrum. Meth. 164 (1979)
  267--278.
\newblock \href {http://dx.doi.org/10.1016/0029-554X(79)90244-1}
  {\path{doi:10.1016/0029-554X(79)90244-1}}.

\bibitem{Learned:1978iv}
J.~G. Learned, {Acoustic Radiation by Charged Atomic Particles in Liquids: An
  Analysis}, Phys. Rev. D 19 (1979) 3293.
\newblock \href {http://dx.doi.org/10.1103/PhysRevD.19.3293}
  {\path{doi:10.1103/PhysRevD.19.3293}}.

\bibitem{Abdou:2011cy}
Y.~Abdou, et~al., {Design and performance of the South Pole Acoustic Test
  Setup}, Nucl. Instrum. Meth. A 683 (2012) 78--90.
\newblock \href {http://arxiv.org/abs/1105.4339} {\path{arXiv:1105.4339}},
  \href {http://dx.doi.org/10.1016/j.nima.2012.03.027}
  {\path{doi:10.1016/j.nima.2012.03.027}}.

\bibitem{Riccobene:2009zz}
G.~Riccobene, et~al., {Long-term measurements of acoustic background noise in
  very deep sea}, Nucl. Instrum. Meth. A 604 (2009) S149--S157.
\newblock \href {http://dx.doi.org/10.1016/j.nima.2009.03.195}
  {\path{doi:10.1016/j.nima.2009.03.195}}.

\bibitem{Danaher:2007zz}
S.~Danaher, {First data from ACoRNE and signal processing techniques}, J. Phys.
  Conf. Ser. 81 (2007) 012011.
\newblock \href {http://dx.doi.org/10.1088/1742-6596/81/1/012011}
  {\path{doi:10.1088/1742-6596/81/1/012011}}.

\bibitem{Aguilar:2010ac}
J.~Aguilar, et~al., {AMADEUS - The Acoustic Neutrino Detection Test System of
  the ANTARES Deep-Sea Neutrino Telescope}, Nucl. Instrum. Meth. A 626-627
  (2011) 128--143.
\newblock \href {http://arxiv.org/abs/1009.4179} {\path{arXiv:1009.4179}},
  \href {http://dx.doi.org/10.1016/j.nima.2010.09.053}
  {\path{doi:10.1016/j.nima.2010.09.053}}.

\bibitem{Vandenbroucke:2004gv}
J.~Vandenbroucke, G.~Gratta, N.~Lehtinen, {Experimental study of acoustic
  ultrahigh - energy neutrino detection}, Astrophys. J. 621 (2005) 301--312.
\newblock \href {http://arxiv.org/abs/astro-ph/0406105}
  {\path{arXiv:astro-ph/0406105}}, \href {http://dx.doi.org/10.1086/425336}
  {\path{doi:10.1086/425336}}.

\bibitem{2017EPJWC.13506002S}
F.~{Simeone}, A.~{Capone}, {Acoustic detection of UHE neutrinos in the
  Mediterranean sea: status and perspective}, in: European Physical Journal Web
  of Conferences, Vol. 135 of European Physical Journal Web of Conferences,
  2017, p. 06002.
\newblock \href {http://dx.doi.org/10.1051/epjconf/201713506002}
  {\path{doi:10.1051/epjconf/201713506002}}.

\bibitem{IceCube-Gen2:2020qha}
M.~G. Aartsen, et~al., {IceCube-Gen2: the window to the extreme Universe}, J.
  Phys. G 48~(6) (2021) 060501.
\newblock \href {http://arxiv.org/abs/2008.04323} {\path{arXiv:2008.04323}},
  \href {http://dx.doi.org/10.1088/1361-6471/abbd48}
  {\path{doi:10.1088/1361-6471/abbd48}}.

\bibitem{Berezinsky:2002nc}
V.~Berezinsky, A.~Z. Gazizov, S.~I. Grigorieva, {On astrophysical solution to
  ultrahigh-energy cosmic rays}, Phys. Rev. D 74 (2006) 043005.
\newblock \href {http://arxiv.org/abs/hep-ph/0204357}
  {\path{arXiv:hep-ph/0204357}}, \href
  {http://dx.doi.org/10.1103/PhysRevD.74.043005}
  {\path{doi:10.1103/PhysRevD.74.043005}}.

\bibitem{Aloisio:2012ba}
R.~Aloisio, V.~Berezinsky, A.~Gazizov, {Transition from galactic to
  extragalactic cosmic rays}, Astropart. Phys. 39-40 (2012) 129--143.
\newblock \href {http://arxiv.org/abs/1211.0494} {\path{arXiv:1211.0494}},
  \href {http://dx.doi.org/10.1016/j.astropartphys.2012.09.007}
  {\path{doi:10.1016/j.astropartphys.2012.09.007}}.

\bibitem{deBernardis:2000sbo}
P.~de~Bernardis, et~al., {A Flat universe from high resolution maps of the
  cosmic microwave background radiation}, Nature 404 (2000) 955--959.
\newblock \href {http://arxiv.org/abs/astro-ph/0004404}
  {\path{arXiv:astro-ph/0004404}}, \href {http://dx.doi.org/10.1038/35010035}
  {\path{doi:10.1038/35010035}}.

\bibitem{Franceschini:2008tp}
A.~Franceschini, G.~Rodighiero, M.~Vaccari, {The extragalactic optical-infrared
  background radiations, their time evolution and the cosmic photon-photon
  opacity}, Astron. Astrophys. 487 (2008) 837.
\newblock \href {http://arxiv.org/abs/0805.1841} {\path{arXiv:0805.1841}},
  \href {http://dx.doi.org/10.1051/0004-6361:200809691}
  {\path{doi:10.1051/0004-6361:200809691}}.

\bibitem{Mucke:1998mk}
A.~Mucke, J.~P. Rachen, R.~Engel, R.~J. Protheroe, T.~Stanev, {On photohadronic
  processes in astrophysical environments}, Publ. Astron. Soc. Austral. 16
  (1999) 160.
\newblock \href {http://arxiv.org/abs/astro-ph/9808279}
  {\path{arXiv:astro-ph/9808279}}, \href {http://dx.doi.org/10.1071/AS99160}
  {\path{doi:10.1071/AS99160}}.

\bibitem{Greisen:1966jv}
K.~Greisen, {End to the cosmic ray spectrum?}, Phys. Rev. Lett. 16 (1966)
  748--750.
\newblock \href {http://dx.doi.org/10.1103/PhysRevLett.16.748}
  {\path{doi:10.1103/PhysRevLett.16.748}}.

\bibitem{Zatsepin:1966jv}
G.~T. Zatsepin, V.~A. Kuzmin, {Upper limit of the spectrum of cosmic rays},
  JETP Lett. 4 (1966) 78--80.

\bibitem{Das:2020nvx}
S.~Das, S.~Razzaque, N.~Gupta, {Modeling the spectrum and composition of
  ultrahigh-energy cosmic rays with two populations of extragalactic sources},
  Eur. Phys. J. C 81~(1) (2021) 59.
\newblock \href {http://arxiv.org/abs/2004.07621} {\path{arXiv:2004.07621}},
  \href {http://dx.doi.org/10.1140/epjc/s10052-021-08885-4}
  {\path{doi:10.1140/epjc/s10052-021-08885-4}}.

\bibitem{Ackermann:2014usa}
M.~Ackermann, et~al., {The spectrum of isotropic diffuse gamma-ray emission
  between 100 MeV and 820 GeV}, Astrophys. J. 799 (2015) 86.
\newblock \href {http://arxiv.org/abs/1410.3696} {\path{arXiv:1410.3696}},
  \href {http://dx.doi.org/10.1088/0004-637X/799/1/86}
  {\path{doi:10.1088/0004-637X/799/1/86}}.

\bibitem{Hillas:1984ijl}
A.~M. Hillas, {The Origin of Ultrahigh-Energy Cosmic Rays}, Ann. Rev. Astron.
  Astrophys. 22 (1984) 425--444.
\newblock \href {http://dx.doi.org/10.1146/annurev.aa.22.090184.002233}
  {\path{doi:10.1146/annurev.aa.22.090184.002233}}.

\bibitem{Waxman:1997ti}
E.~Waxman, J.~N. Bahcall, {High-energy neutrinos from cosmological gamma-ray
  burst fireballs}, Phys. Rev. Lett. 78 (1997) 2292--2295.
\newblock \href {http://arxiv.org/abs/astro-ph/9701231}
  {\path{arXiv:astro-ph/9701231}}, \href
  {http://dx.doi.org/10.1103/PhysRevLett.78.2292}
  {\path{doi:10.1103/PhysRevLett.78.2292}}.

\bibitem{Urry:1995mg}
C.~M. Urry, P.~Padovani, {Unified schemes for radio-loud active galactic
  nuclei}, Publ. Astron. Soc. Pac. 107 (1995) 803.
\newblock \href {http://arxiv.org/abs/astro-ph/9506063}
  {\path{arXiv:astro-ph/9506063}}, \href {http://dx.doi.org/10.1086/133630}
  {\path{doi:10.1086/133630}}.

\bibitem{Berezinsky:1996wx}
V.~S. Berezinsky, P.~Blasi, V.~S. Ptuskin, {Clusters of Galaxies as a Storage
  Room for Cosmic Rays}, Astrophys. J. 487 (1997) 529--535.
\newblock \href {http://arxiv.org/abs/astro-ph/9609048}
  {\path{arXiv:astro-ph/9609048}}, \href {http://dx.doi.org/10.1086/304622}
  {\path{doi:10.1086/304622}}.

\bibitem{IceCube:2018dnn}
M.~G. Aartsen, et~al., {Multimessenger observations of a flaring blazar
  coincident with high-energy neutrino IceCube-170922A}, Science 361~(6398)
  (2018) eaat1378.
\newblock \href {http://arxiv.org/abs/1807.08816} {\path{arXiv:1807.08816}},
  \href {http://dx.doi.org/10.1126/science.aat1378}
  {\path{doi:10.1126/science.aat1378}}.

\bibitem{IceCube:2018cha}
M.~G. Aartsen, et~al., {Neutrino emission from the direction of the blazar TXS
  0506+056 prior to the IceCube-170922A alert}, Science 361~(6398) (2018)
  147--151.
\newblock \href {http://arxiv.org/abs/1807.08794} {\path{arXiv:1807.08794}},
  \href {http://dx.doi.org/10.1126/science.aat2890}
  {\path{doi:10.1126/science.aat2890}}.

\bibitem{Paiano:2018qeq}
S.~Paiano, R.~Falomo, A.~Treves, R.~Scarpa, {The redshift of the BL Lac object
  TXS 0506+056}, Astrophys. J. Lett. 854~(2) (2018) L32.
\newblock \href {http://arxiv.org/abs/1802.01939} {\path{arXiv:1802.01939}},
  \href {http://dx.doi.org/10.3847/2041-8213/aaad5e}
  {\path{doi:10.3847/2041-8213/aaad5e}}.

\bibitem{Murase:2014foa}
K.~Murase, Y.~Inoue, C.~D. Dermer, {Diffuse Neutrino Intensity from the Inner
  Jets of Active Galactic Nuclei: Impacts of External Photon Fields and the
  Blazar Sequence}, Phys. Rev. D 90~(2) (2014) 023007.
\newblock \href {http://arxiv.org/abs/1403.4089} {\path{arXiv:1403.4089}},
  \href {http://dx.doi.org/10.1103/PhysRevD.90.023007}
  {\path{doi:10.1103/PhysRevD.90.023007}}.

\bibitem{Aartsen:2016oji}
M.~G. Aartsen, et~al., {All-sky Search for Time-integrated Neutrino Emission
  from Astrophysical Sources with 7 yr of IceCube Data}, Astrophys. J. 835~(2)
  (2017) 151.
\newblock \href {http://arxiv.org/abs/1609.04981} {\path{arXiv:1609.04981}},
  \href {http://dx.doi.org/10.3847/1538-4357/835/2/151}
  {\path{doi:10.3847/1538-4357/835/2/151}}.

\bibitem{2021MNRAS.tmp.1320M}
A.~{Marinelli}, J.~R. {Sacahui}, A.~{Sharma}, M.~{Osorio-Archila}, {Analyzing
  the gamma-ray activity of neutrino emitter candidates: comparing TXS 0506+056
  with other blazars}, Montly Notices of the Royal Astronomical Society\href
  {http://dx.doi.org/10.1093/mnras/stab1312}
  {\path{doi:10.1093/mnras/stab1312}}.

\bibitem{Oikonomou:2019djc}
F.~Oikonomou, K.~Murase, P.~Padovani, E.~Resconi, P.~M\'esz\'aros, {High energy
  neutrino flux from individual blazar flares}, Mon. Not. Roy. Astron. Soc.
  489~(3) (2019) 4347--4366.
\newblock \href {http://arxiv.org/abs/1906.05302} {\path{arXiv:1906.05302}},
  \href {http://dx.doi.org/10.1093/mnras/stz2246}
  {\path{doi:10.1093/mnras/stz2246}}.

\bibitem{Fraija:2017jok}
N.~Fraija, E.~Aguilar-Ruiz, A.~Galv\'an-G\'amez, A.~Marinelli, J.~A. de~Diego,
  {Study of the PeV neutrino, \ensuremath{\gamma}-rays, and UHECRs around the
  lobes of Centaurus A}, Mon. Not. Roy. Astron. Soc. 481~(4) (2018) 4461--4471.
\newblock \href {http://arxiv.org/abs/1709.05766} {\path{arXiv:1709.05766}},
  \href {http://dx.doi.org/10.1093/mnras/sty2561}
  {\path{doi:10.1093/mnras/sty2561}}.

\bibitem{ThePierreAuger:2015rma}
A.~Aab, et~al., {The Pierre Auger Cosmic Ray Observatory}, Nucl. Instrum. Meth.
  A 798 (2015) 172--213.
\newblock \href {http://arxiv.org/abs/1502.01323} {\path{arXiv:1502.01323}},
  \href {http://dx.doi.org/10.1016/j.nima.2015.06.058}
  {\path{doi:10.1016/j.nima.2015.06.058}}.

\bibitem{Clarke:2004te}
T.~E. Clarke, {Faraday rotation observations of magnetic fields in galaxy
  clusters}, J. Korean Astron. Soc. 37~(5) (2004) 337--342.
\newblock \href {http://arxiv.org/abs/astro-ph/0412268}
  {\path{arXiv:astro-ph/0412268}}, \href
  {http://dx.doi.org/10.5303/JKAS.2004.37.5.337}
  {\path{doi:10.5303/JKAS.2004.37.5.337}}.

\bibitem{Ensslin:2005uk}
T.~A. Ensslin, C.~Vogt, {Magnetic turbulence in cool cores of galaxy clusters},
  Astron. Astrophys. 453 (2006) 447.
\newblock \href {http://arxiv.org/abs/astro-ph/0505517}
  {\path{arXiv:astro-ph/0505517}}, \href
  {http://dx.doi.org/10.1051/0004-6361:20053518}
  {\path{doi:10.1051/0004-6361:20053518}}.

\bibitem{Fang:2013vla}
K.~Fang, K.~Kotera, K.~Murase, A.~V. Olinto, {Testing the Newborn Pulsar Origin
  of Ultrahigh Energy Cosmic Rays with EeV Neutrinos}, Phys. Rev. D 90~(10)
  (2014) 103005, [Erratum: Phys.Rev.D 92, 129901 (2015)].
\newblock \href {http://arxiv.org/abs/1311.2044} {\path{arXiv:1311.2044}},
  \href {http://dx.doi.org/10.1103/PhysRevD.90.103005}
  {\path{doi:10.1103/PhysRevD.90.103005}}.

\bibitem{Matthews:2018rpe}
J.~H. Matthews, A.~R. Bell, K.~M. Blundell, A.~T. Araudo, {Ultrahigh energy
  cosmic rays from shocks in the lobes of powerful radio galaxies}, Mon. Not.
  Roy. Astron. Soc. 482~(4) (2019) 4303--4321.
\newblock \href {http://arxiv.org/abs/1810.12350} {\path{arXiv:1810.12350}},
  \href {http://dx.doi.org/10.1093/mnras/sty2936}
  {\path{doi:10.1093/mnras/sty2936}}.

\bibitem{Griest:1989wd}
K.~Griest, M.~Kamionkowski, {Unitarity Limits on the Mass and Radius of Dark
  Matter Particles}, Phys. Rev. Lett. 64 (1990) 615.
\newblock \href {http://dx.doi.org/10.1103/PhysRevLett.64.615}
  {\path{doi:10.1103/PhysRevLett.64.615}}.

\bibitem{Smirnov:2019ngs}
J.~Smirnov, J.~F. Beacom, {TeV-Scale Thermal WIMPs: Unitarity and its
  Consequences}, Phys. Rev. D 100~(4) (2019) 043029.
\newblock \href {http://arxiv.org/abs/1904.11503} {\path{arXiv:1904.11503}},
  \href {http://dx.doi.org/10.1103/PhysRevD.100.043029}
  {\path{doi:10.1103/PhysRevD.100.043029}}.

\bibitem{Zavala:2014dla}
J.~Zavala, {Galactic PeV neutrinos from dark matter annihilation}, Phys. Rev. D
  89~(12) (2014) 123516.
\newblock \href {http://arxiv.org/abs/1404.2932} {\path{arXiv:1404.2932}},
  \href {http://dx.doi.org/10.1103/PhysRevD.89.123516}
  {\path{doi:10.1103/PhysRevD.89.123516}}.

\bibitem{Chianese:2021htv}
M.~Chianese, D.~F.~G. Fiorillo, R.~Hajjar, G.~Miele, S.~Morisi, N.~Saviano,
  {Heavy decaying dark matter at future neutrino radio telescopes}, JCAP 05
  (2021) 074.
\newblock \href {http://arxiv.org/abs/2103.03254} {\path{arXiv:2103.03254}},
  \href {http://dx.doi.org/10.1088/1475-7516/2021/05/074}
  {\path{doi:10.1088/1475-7516/2021/05/074}}.

\bibitem{misurevelocita}
S.~Salon, A.~Crise, P.~Picco, E.~de~Marinis, O.~Gasparini,
  \href{https://angeo.copernicus.org/articles/21/833/2003/}{Sound speed in the
  mediterranean sea: an analysis from a climatological data set}, Annales
  Geophysicae 21~(3) (2003) 833--846.
\newblock \href {http://dx.doi.org/10.5194/angeo-21-833-2003}
  {\path{doi:10.5194/angeo-21-833-2003}}.
\newline\urlprefix\url{https://angeo.copernicus.org/articles/21/833/2003/}

\bibitem{https://doi.org/10.2312/gfz.nmsop-2ch9}
J.~Schweitzer, J.~Fyen, S.~Mykkeltveit, S.~J. Gibbons, M.~Pirli, D.~K\"{u}hn,
  T.~Kværna,
  \href{https://gfzpublic.gfz-potsdam.de/pubman/item/item_43213}{Seismic
  arrays}, New Manual of Seismological Observatory Practice 2 (NMSOP2)\href
  {http://dx.doi.org/10.2312/GFZ.NMSOP-2_CH9}
  {\path{doi:10.2312/GFZ.NMSOP-2_CH9}}.
\newline\urlprefix\url{https://gfzpublic.gfz-potsdam.de/pubman/item/item_43213}

\bibitem{Niess2006}
V.~Niess, V.~Bertin,
  \href{https://doi.org/10.1016/j.astropartphys.2006.06.005}{Underwater
  acoustic detection of ultra high energy neutrinos}, Astroparticle Physics
  26~(4-5) (2006) 243--256.
\newblock \href {http://dx.doi.org/10.1016/j.astropartphys.2006.06.005}
  {\path{doi:10.1016/j.astropartphys.2006.06.005}}.
\newline\urlprefix\url{https://doi.org/10.1016/j.astropartphys.2006.06.005}

\bibitem{BEVAN2009398}
S.~Bevan, A.~Brown, S.~Danaher, J.~Perkin, C.~Rhodes, T.~Sloan, L.~Thompson,
  O.~Veledar, D.~Waters,
  \href{https://www.sciencedirect.com/science/article/pii/S0168900209009401}{Study
  of the acoustic signature of uhe neutrino interactions in water and ice},
  Nuclear Instruments and Methods in Physics Research Section A: Accelerators,
  Spectrometers, Detectors and Associated Equipment 607~(2) (2009) 398--411.
\newblock \href {http://dx.doi.org/https://doi.org/10.1016/j.nima.2009.05.009}
  {\path{doi:https://doi.org/10.1016/j.nima.2009.05.009}}.
\newline\urlprefix\url{https://www.sciencedirect.com/science/article/pii/S0168900209009401}

\bibitem{idrofono}
B.~. K. S. .~V. Measurement,
  \href{https://www.bksv.com/en/transducers/acoustic/microphones/hydrophones/8106}{Type
  8106 low-noise hydrophone with preamplifier}, in: Modeling and Measurement
  Methods for Acoustic Waves and for Acoustic Microdevices, {InTech}, 2021.
\newline\urlprefix\url{https://www.bksv.com/en/transducers/acoustic/microphones/hydrophones/8106}

\bibitem{libroneacustica}
L.~E. {Kinsler}, A.~R. {Frey}, A.~B. {Coppens}, J.~V. {Sanders}, {Fundamentals
  of Acoustics, 4th Edition}, 1999.

\bibitem{Kuperman2004}
W.~A. Kuperman, J.~F. Lynch,
  \href{https://doi.org/10.1063/1.1825269}{Shallow-water acoustics}, Physics
  Today 57~(10) (2004) 55--61.
\newblock \href {http://dx.doi.org/10.1063/1.1825269}
  {\path{doi:10.1063/1.1825269}}.
\newline\urlprefix\url{https://doi.org/10.1063/1.1825269}

\bibitem{SousaCosta2013}
E.~de~Sousa~Costa, E.~Bauzer, J.~B.~C. Filardi,
  \href{https://doi.org/10.5772/56132}{Underwater acoustics modeling in finite
  depth shallow waters}, in: Modeling and Measurement Methods for Acoustic
  Waves and for Acoustic Microdevices, {InTech}, 2013.
\newblock \href {http://dx.doi.org/10.5772/56132} {\path{doi:10.5772/56132}}.
\newline\urlprefix\url{https://doi.org/10.5772/56132}

\bibitem{Chen2016}
F.~Chen, Y.~Li, L.~Lv, \href{https://doi.org/10.1109/coa.2016.7535676}{Analysis
  of shallow water sound velocity profile impact on detection performance of
  active sonar signal}, in: 2016 {IEEE}/{OES} China Ocean Acoustics ({COA}),
  {IEEE}, 2016.
\newblock \href {http://dx.doi.org/10.1109/coa.2016.7535676}
  {\path{doi:10.1109/coa.2016.7535676}}.
\newline\urlprefix\url{https://doi.org/10.1109/coa.2016.7535676}

\bibitem{Porter2010}
M.~B. Porter, M.~Siderius,
  \href{https://doi.org/10.1109/wssc.2010.5730274}{Acoustic propagation in very
  shallow water}, in: 2010 International {WaterSide} Security Conference,
  {IEEE}, 2010.
\newblock \href {http://dx.doi.org/10.1109/wssc.2010.5730274}
  {\path{doi:10.1109/wssc.2010.5730274}}.
\newline\urlprefix\url{https://doi.org/10.1109/wssc.2010.5730274}

\bibitem{wenz}
G.~M. Wenz, \href{https://doi.org/10.1121/1.1909155}{Acoustic ambient noise in
  the ocean: Spectra and sources}, The Journal of the Acoustical Society of
  America 34~(12) (1962) 1936--1956.
\newblock \href {http://arxiv.org/abs/https://doi.org/10.1121/1.1909155}
  {\path{arXiv:https://doi.org/10.1121/1.1909155}}, \href
  {http://dx.doi.org/10.1121/1.1909155} {\path{doi:10.1121/1.1909155}}.
\newline\urlprefix\url{https://doi.org/10.1121/1.1909155}

\bibitem{Cooper-Sarkar:2007zsa}
A.~Cooper-Sarkar, S.~Sarkar, {Predictions for high energy neutrino
  cross-sections from the ZEUS global PDF fits}, JHEP 01 (2008) 075.
\newblock \href {http://arxiv.org/abs/0710.5303} {\path{arXiv:0710.5303}},
  \href {http://dx.doi.org/10.1088/1126-6708/2008/01/075}
  {\path{doi:10.1088/1126-6708/2008/01/075}}.

\end{thebibliography}
\end{document}